\documentclass[fleqn,usenatbib]{mnras}

\usepackage[utf8]{inputenc}
\usepackage[T1]{fontenc}
\usepackage{ae,aecompl}
\usepackage{graphicx}
\graphicspath{{./}}
\usepackage{amsmath}
\usepackage{amssymb}	
\usepackage{listings}
\usepackage{subcaption}
\usepackage{longtable}
\usepackage[justification=justified, singlelinecheck=false]{caption}
\usepackage{hyperref}
\usepackage{natbib}
\usepackage[pagewise,left]{lineno}

\newcommand{\gaia}{\textit{Gaia}}
\newcommand{\gdrthree}{\textit{Gaia}~DR3}
\newcommand{\espucd}{ESP-UCD}
\newcommand{\flagsespucd}{\texttt{flags\_espucd}}
\newcommand{\teffespucd}{\texttt{teff\_espucd}}
\newcommand{\crall}{\ensuremath{\texttt{CR}}}
\newcommand{\teff}{\ensuremath{\mathrm{T_{eff}}}}
\newcommand{\colour}{\texttt{colour}}
\newcommand{\blue}{\texttt{blue}}
\newcommand{\red}{\texttt{red}}
\newcommand{\kms}{\ensuremath{\mathrm{km\,s}^{-1}}}
\newcommand{\vtotal}{\ensuremath{V_{\text{total}}}}
\newcommand{\vtan}{\ensuremath{V_{\text{tan}}}}
\newcommand{\phalo}{\ensuremath{P_{\text{halo}}}}
\newcommand{\pthick}{\ensuremath{P_{\text{thick}}}}
\newcommand{\pthin}{\ensuremath{P_{\text{thin}}}}
\newcommand{\ntotal}{\ensuremath{21\,205}}
\newcommand{\ngucdscross}{\ensuremath{2565}}
\newcommand{\noutlier}{\ensuremath{58}}
\newcommand{\nphotosd}{\ensuremath{260}}
\newcommand{\nphotoyo}{\ensuremath{906}}
\newcommand{\nhalocand}{\ensuremath{27}}
\newcommand{\nthickcand}{\ensuremath{3701}}
\newcommand{\noutphotsd}{six}
\newcommand{\noutphotyo}{one}
\newcommand{\nsdcand}{six}
\newcommand{\nyocand}{one}
\newcommand{\nprimecand}{seven}

\title[\textbf{Ultracool Outliers in Gaia DR3}]{\textbf{Ultracool Spectroscopic Outliers in \gaia\ DR3}}
\author[W.\,J. Cooper et~al.]{W.\,J.~Cooper,$^{1,2}$\thanks{E-mail: w.cooper@herts.ac.uk}
R.\,L.~Smart,$^{2}$
H.\,R.\,A.~Jones,$^{1}$
L.\,M.~Sarro$^{3}$\\
$^{1}$Centre for Astrophysics Research, University of Hertfordshire, Hatfield, Hertfordshire, AL10 9AB, UK\\
$^{2}$Istituto Nazionale di Astrofisica, Osservatorio Astrofisico di Torino, Strada Osservatorio 20, I-10025 Pino Torinese, IT\\
$^{3}$Departamento de Inteligencia Artificial, ETSI Informática, UNED, Juan del Rosal, E-16 28040 Madrid, ES
}
\date{Accepted 2023 October 02. Received 2023 September 28; in original form 2023 July 24}
\pubyear{2023} 

\begin{document}
\label{firstpage}
\pagerange{\pageref{firstpage}--\pageref{lastpage}}
\maketitle

\begin{abstract}
    \gdrthree\ provided a first release of RP spectra and astrophysical parameters for ultracool dwarfs.
    We used these \gaia\ RP spectra and astrophysical parameters to select the most outlying ultracool dwarfs.
    These objects have spectral types of M7 or later and might be young brown dwarfs
    or low metallicity objects.
    This work aimed to find ultracool dwarfs which have \gaia\ RP spectra
    significantly different to the typical population.
    However, the intrinsic faintness of these ultracool dwarfs in \gaia\ means that their
    spectra were typically rather low signal-to-noise in \gdrthree\@.
    This study is intended as a proof-of-concept for future iterations of the \gaia\ data releases.
    Based on well studied subdwarfs and young objects, we created a spectral type-specific color ratio,
    defined using \gaia\ RP spectra;
    this ratio is then used to determine which objects are outliers.
    We then used the objects kinematics and photometry external to \gaia\ to cut down
    the list of outliers into a list of ‘prime candidates’.
    We produce a list of \noutlier\ \gaia\ RP spectra outliers, \nprimecand\ of which we deem as prime candidates.
    Of these, \nsdcand\ are likely subdwarfs and \nyocand\ is a known young stellar object.
    Four of \nsdcand\ subdwarf candidates
    were known as subdwarfs already.
    The two other subdwarf candidates: 2MASS J03405673$+$2633447 (sdM8.5) and
    2MASS J01204397$+$6623543 (sdM9), are new classifications.
\end{abstract}


\section{Introduction} \label{sec:dr3intro}
Ultracool dwarfs (UCDs) are objects with spectral types cooler than M7 ($\teff \lessapprox 2700$\,K),
consisting of late M, L, T and Y dwarfs.
These newest spectral types were first described by~\citet{1999ApJ...519..802K},
\citet{2002ApJ...564..421B}and~\citet{2011ApJ...743...50C}.
Spectral types of UCDs are primarily driven by changes in effective temperature,
while other features (e.g., low-surface gravity, low-metallicity)
can further refine them~\citep[see][]{2005ARA&A..43..195K}.
The aim of this work is to use the Gaia data to select outlying UCDs and in particular,
the youngest and oldest examples.

Subdwarfs are old objects, with lower metallicities than field objects.
As such, multi-wavelength photometric cross-matches are an ideal method to select subdwarf candidates.
Notably, optical surveys like \gaia~\citep{2016A&A...595A...1G} and Pan-STARRS~\citep{2016arXiv161205560C} are
typically compared with near/mid-infrared surveys including 2MASS~\citep{2006AJ....131.1163S} and
AllWISE~\citep{2014yCat.2328....0C}.
Kinematically, subdwarfs, due to their age, are much faster than field objects.
Hence, subdwarfs (depending on their metallicity and age) are either thick disk or halo objects.
Multiple literature sources discuss the selections and classifications of thick disk/halo dwarfs,
for example, work by~\citet{1992ApJS...82..351L}.
For purely kinematic selections of halo objects, when metallicity information is not present,
~\citet{2010A&A...511L..10N} utilised either a cut of $\vtotal > 180$\,\kms~\citep{2004AJ....128.1177V} or
$\vtotal > 210$\,\kms~\citep[][depending on the Galactic model used]{2009MNRAS.399.1145S, 2018ApJ...860L..11K}
where $\vtotal$ is the total space velocity, $\vtotal = \sqrt {U^2 + V^2 + W^2}$, and $U, V, W$
are the velocities in the Galactic reference frame.
Likewise, selection of thick disk objects varies from $\vtotal > 85$\,\kms~\citep{2006A&A...449..127Z} to
$\vtotal > 70$\,\kms~\citep{2010A&A...511L..10N} and $\vtotal > 50$\,\kms~\citep{2023A&A...674A..39G}.
Without radial velocity (RV) information, tangential velocity, \vtan, has been often used as it is highly
indicative of thick disc/halo membership.
Ultracool subdwarfs follow this same detection criteria~\citep{1997AJ....113..806G, 1999AJ....117..508G}.
We follow previous work discovering ultracool subdwarfs~\citep[e.g.,][]{2017MNRAS.468..261Z, 2019MNRAS.486.1840Z}
which has benefit from the selection of subdwarfs using virtual observatory
tools~\citep{2012A&A...542A.105L, 2017A&A...598A..92L} and all-sky
surveys~\citep{2002AJ....124.1190L, 2008AJ....135.2177L}.

By comparison, young objects have typically lower surface
gravities and are redder than field objects~\citep{2016AAS...22714503C}.
Unresolved binaries often occupy the same space on colour-absolute magnitude diagrams (CMDs) as young objects,
hence purely photometric selections are contaminated~\citep[e.g.,][]{2017MNRAS.470.4885M}.
Kinematically, young objects are slower than field objects, and are often still gravitationally bound
to young moving groups~\citep[and references therein]{2018ApJ...862..138G}.
Gathering spectra of UCD candidates is therefore necessary for confirming youth,
especially when the objects are isolated.
The spectral confirmation of youth involves analysing the surface gravity of the UCD, where a lower gravity indicates
a younger object.
Optical spectra are given Greek letter classifications
with $\alpha$ as normal, $\beta$ as intermediate, $\gamma$ as low gravity~\citep{2009AJ....137.3345C} and $\delta$ for extreme low
gravity~\citep{2006ApJ...639.1120K}.

\gaia\ is a European Space Agency mission launched in 2013 and in June 2022 released
\gdrthree~\citep{2023A&A...674A...1G} which, importantly for this work, included spectra.
This is referred to as `XP' spectra where `X' can be interchanged with either `B' or `R' corresponding
to the blue and red filters.
\gaia\ provides five dimensional astrometric measurements (two positions, two proper motions and parallax).
\gaia\ also released RVs for objects with $G_{\mathrm{RVS}} \lessapprox 14$\,mag~\citep{2023A&A...674A...5K},
where $G_{\mathrm{RVS}}$ is the magnitude integrated across the \gaia\ RV
spectrometer~\citep[RVS,][]{2023A&A...674A...6S}.
We focus here on RP spectra, which cover the far red optical regime from ${\approx}600\text{--}1050$\,nm.
The resolution of these internally calibrated spectra for UCDs are around
$30\text{--}50\,\frac{\Delta \lambda}{\lambda}$~\citep[][who also discuss
the external calibration]{2023A&A...674A...3M}.

\gaia\ is well-suited to observe nearby early-type
UCDs~\citep[see fig.~26,][${<}$L5, $d < 30$\,pc]{2021A&A...649A...6G}.
Known \gaia\ UCDs are documented in the \gaia\ Ultracool Dwarf
Sample~\citep[GUCDS -][Cooper et~al. \textit{submitted}]{2017MNRAS.469..401S, 2019MNRAS.485.4423S, 2020MNRAS.494.4891M}.
GUCDS aims to be complete for known L dwarfs but also contains late-M dwarfs, T dwarfs and
primary stars from any relevant common proper motion systems.
Volume limited samples have been vital for understanding the UCD population,
as performed by~\citet{2021A&A...649A...6G}, \citet{2021ApJS..253....7K} and~\citet{2021A&A...650A.201R}.
We focus on late-M to L dwarfs, for which the spectral features evolve as described by~\citet{1998MNRAS.301.1031T}
and~\citet{1999ApJ...519..802K}.
However, at the low resolution of \gaia\ RP spectra, individual features cannot be seen,
leading to a merging of features~\citep{2023A&A...669A.139S}.

Recently, many discoveries have been using \gaia\ data with the focus of finding outlying objects
and astrophysical parameters.
For example, exploration of hot subdwarf stars in \gdrthree~\citep{2022A&A...662A..40C}
found 21\,785 underluminous objects.
\citet{2023arXiv230317676Y} uncovered 188\,000 candidate metal-poor stars using \gaia\ XP spectra.
Similarly, \citet{2023ApJS..267....8A}, following the study by~\citet{2023arXiv230206995A},
applied \texttt{XGBoost} to determine metallicities for main-sequence dwarfs and giants.
Parameters of stars, forward modelled from \gaia\ XP spectra, were also determined by~\citet{2023MNRAS.tmp.1869Z}.

\begin{figure}
    \includegraphics[width=\linewidth]{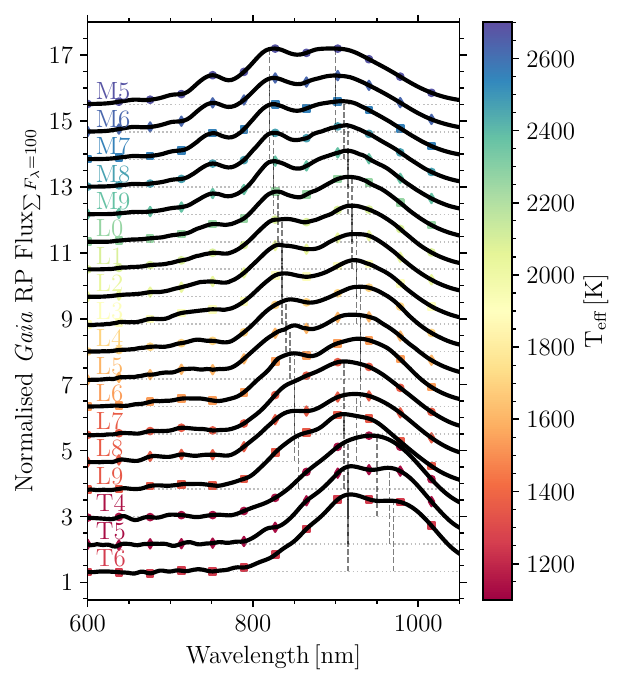}
    \caption{
        The normalised median RP fluxes for each spectral type (see Sect.~\ref{subsec:spectyping}) from M5--T6.
        Each spectral type is indicated by the attached text with its corresponding median effective temperature
        given on the auxiliary axis.
        Vertical dashed lines are shown for every spectrum to indicate the position of the two primary spectral peaks.
        The normalised spectra were multiplied by a constant value such that the fluxes sum to 100 instead of 1
        and are offset by a set value.
    }
    \label{fig:gucdsmedians}
\end{figure}

In UCDs, spectral feature changes due to age or metallicity are not directly seen in the RP spectra,
as the spectra are too low resolution to readily
be isolated, they do however change the general shape of the RP spectra,
most notably the centroids and intensity of the
2--3 peaks~\citep[Fig.~\ref{fig:gucdsmedians} in this work and fig.~5 by][]{2023A&A...669A.139S}.
As effective temperature decreases in Fig.~\ref{fig:gucdsmedians},
the first peak (${\sim}750$\,nm) disappears when approaching the stellar/substellar
boundary~\citep[${\approx}\mathrm{L}3$,][]{2021A&A...649A...6G}
whilst the second peak goes from being brighter than the third peak in M dwarfs, to being dimmer than the third peak
in L dwarfs and being roughly equivalent in T dwarfs.
In addition, the centroids of the peaks redshift with decreasing \teff.

This work is focused on the characterisation of the \gaia\ internally calibrated RP spectra
and the isolation of young and subdwarf UCDs.
Section~\ref{sec:dr3method} discusses the methodology, and the creation of a colour ratio;
Section~\ref{sec:dr3analysis} is the analysis and selection of prime candidates from external photometry and kinematics;
Section~\ref{sec:dr3results} show the results of our prime candidates whilst
Section~\ref{sec:dr3discussion} concludes and plans future work to counter the known issues.

\section{Method} \label{sec:dr3method}
Here we discuss our iterative approach to deriving an outlier classifier.
We started with the sample of UCDs in \gaia\, as discussed by~\citet{2023A&A...669A.139S}.
To summarise, the sample of \gaia\ UCDs consists of every object for which the \espucd\ work package derived
an effective temperature.
The selection of UCDs from \gaia\ was:
$\varpi > 1.7$\,mas, $G-G_{\text{RP}} > 1$,
$q_{33} > 60$, $q_{50} > 71$, $q_{67} > 83$ where $q_{33},\ q_{50},\ q_{67}$
are the 33.33, 50, 66.67 percentiles of the total RP flux respectively~\citep{2023A&A...674A..26C}.
Of these 94\,158 objects, only \ntotal\ have public RP spectra~\citep[see the
\href{https://gea.esac.esa.int/archive/documentation/GDR3/index.html}{online documentation} and
sect.~4 by][for the \gaia\ spectra publication criteria]{2023A&A...674A...2D}.
All effective temperatures discussed were from \gdrthree, from the astrophysical parameters table and specific to
the UCD work package \espucd.
The relevant columns originating from the \espucd\ work package are \teffespucd\ and \flagsespucd.
One part of the \gdrthree\ RP spectra publication criteria, important for the search of spectral outliers,
was that the \gaia\ RP UCD spectra were required to be one of the highest two quality flags
(0--1, not 2 in \flagsespucd).
The flagging in \espucd\ included measuring the Euclidean distance of a \gaia\ RP spectrum from its BT-Settl
model counterpart.
Whilst this requirement was vital for reducing the number of published \gaia\ RP contaminants, it prejudices
our results against classifying the most extreme spectral outliers, as was expected for extreme and ultra-subdwarfs.
Thus, our expected number of `prime candidates' was diminished.

The RP spectra of these \ntotal\ objects were extracted with 
\texttt{gaiaxpy.convert}~\citep{daniela_ruz_mieres_2022_6674521} through the 
\texttt{gaiaxpy-batch} package~\citep{cooper_w_j_2022_6653446}.
The absolute sampling of the retrieved spectra is a linearly dispersed grid from 600--1050\,nm.
We used this wavelength sampling (and only plot RP spectra within that limit) because it roughly corresponds to
the \gdrthree\ RP passband~\citep[${\approx}620\text{--}1042$\,nm,][]{2021A&A...649A...3R}.
All spectra were divided by the sum of the fluxes across the entire 600--1050\,nm region 
(i.e.\ the total flux of normalised spectra is unity).
This method of normalisation was chosen because other methods (e.g.\ dividing by a median flux of a given
wavelength regime) could cause non-physical artifacts, especially for noisy spectra.
Some \gaia\ RP spectra can exhibit \textit{apparent} negative fluxes, as a result of the projection
onto the Hermite base functions during their construction.
We sample the wavelengths with a consistent linearly dispersed grid.
Ergo, when one normalises all of the \gaia\ RP spectra by dividing by the sum of the fluxes,
the spectra are homogeneous in wavelength and absolute flux calibration, thus are comparable.

Instead of using an absolute magnitude to find outliers, such as the robust $M_{G}$ to spectral type relation,
the \gdrthree\ RP spectral sequence follows the optical spectral features which define spectral
sub-types for different UCDs.
Additionally, as discussed by~\citet{2021A&A...649A...6G}, there is a large scatter in \gaia\ colours for UCDs
for every spectral type bin.
This scatter, present in all photometric selections, means the introduction of a large number of contaminants.
Using spectra instead might prove a cleaner selection technique, even at the low resolution of \gdrthree\ RP spectra.

In this section, we discuss the additional data gathering used to complement \gdrthree\@.
This includes the cross-matching with external photometry as well our basic spectral typing method.
The external photometry was used for validation in Section.~\ref{sec:dr3analysis} whilst the spectral typing
was used to define bins when searching for outliers.
We defined a new colour ratio and used this colour ratio to separate outlying UCDs from normal UCDs.

\subsection{External cross-matching} \label{subsec:crossmatch}
\begin{figure}
    \includegraphics[width=\linewidth]{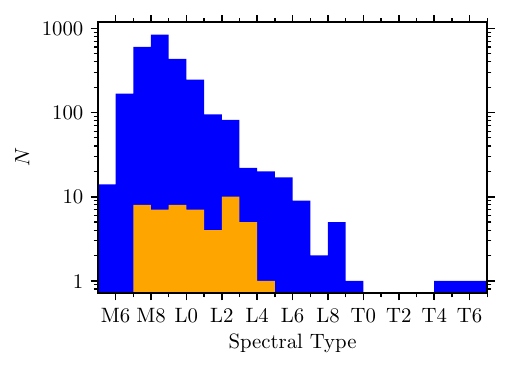}
    \caption{
        Histogram of the number of objects in each spectral type bin from the GUCDS\@.
        The full GUCDS is shown in blue whilst over plotted in orange is the distribution of the known
        standards used.
    }
    \label{fig:gucdshist}
\end{figure}

Using the \gaia\ \href{https://gea.esac.esa.int/archive/}{data archive}, we first performed a
\texttt{`left join'} query against the pre-computed cross-matches for
Pan-STARRS~\citep{2016arXiv161205560C}, 2MASS~\citep{2006AJ....131.1163S}
and AllWISE~\citep{2014yCat.2328....0C}.
From these cross-matches we noted that the Pan-STARRS join was much less complete than 2MASS or AllWISE\@.
As such, Pan-STARRS was not used in the photometric analysis
but was used for the further discussion on our prime candidates.
The RP spectral sample was cross-matched with the GUCDS\@.
The GUCDS contains thousands of known objects with spectral types from the literature.
Of these, ${\approx}270$ are known subdwarfs, and are flagged as such within their spectral types.
This cross-matched sample between our \ntotal\ sample, and the GUCDS, is of size \ngucdscross.
The known subdwarfs and young objects from this GUCDS cross-match are shown in Appendix Table~\ref{tab:litobjects}
and were converted into Boolean flags from which we trained our candidate flagging techniques discussed below.
Additionally, there exists a list of optical standards for a range of
spectral types~\citep[see table~1,][]{2023A&A...669A.139S},
which we use as part of our method and analysis.
This list of standards was supplemented with tens of visually selected bright RP spectra
which were as similar as possible to each standard;
the final list is hereafter referred to as `known standards' and shown in orange in Fig.~\ref{fig:gucdshist}.

\subsection{Estimating a spectral type} \label{subsec:spectyping}
\begin{figure}
    \includegraphics[width=\linewidth]{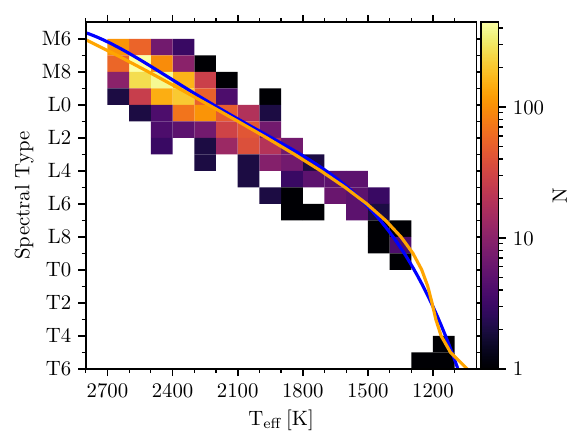}
    \caption{
        Spectral type conversion from \teff\,[K] to spectral type for the GUCDS, as a 2-D histogram.
        The number of objects in each bin is shown by the colour bar.
        Our fourth order polynomial is shown as the blue line.
        By comparison, we plot in orange the fifth order polynomial~\citep[equation~(4):][]{2009ApJ...702..154S}
        relation, valid from M6--T8.
        A wider spread of \teff\ can be seen in the late M and early L dwarfs.
        This is a natural spread as each known spectral type will have an error margin of 1--2 spectral types.
    }
    \label{fig:sptconversion}
\end{figure}

For discussing our objects on an individual basis, it is more meaningful to write in terms of spectral type
than \teff.
As such, we discuss here a simplistic method for estimating spectral type from the \teff\ values
provided by \gdrthree, \teffespucd.
These spectral types estimated here were not used for any analysis.
To more correctly ascertain spectral types, one would match the features and shapes of the RP spectra to well-known
standards.
This, however, is similar to our outlier detection technique, hence we seek to avoid any `cyclic' analysis.
All sources in our RP spectral sample have a derived effective temperature from \gdrthree\@.
However, known objects, including subdwarfs and young objects,
are defined by their spectral types (`SpT', as that is a direct measurement)
rather than effective temperatures, which are generally inferred from modelling.
In the case of \gdrthree, this modelling was trained on an empirical sample not containing any abnormal objects,
like subdwarfs and young objects~\citep{2023A&A...674A..26C, 2023A&A...669A.139S}.
Spectral type is known to have a direct relation to effective temperature, although there is significant scatter
in \teff\ for every spectral type.
To convert the \gaia\ \teffespucd\ into a spectral type we derived a fourth order polynomial between the
\gaia\ \teffespucd\ values and the GUCDS optical spectral types.
This is shown in Fig.~\ref{fig:sptconversion}.
This polynomial follows equation~(\ref{eq:sptpoly}) with coefficients from Table~\ref{tab:polyorders},
where spectral types are converted to numerical values using a code whereby M0=60, L0=70, T0=80, etc.

\begin{equation}
    \label{eq:sptpoly} \mathrm{SpT} = a\teff^4 - b\teff^3 +
    c\teff^2 - d\teff + e
\end{equation}

\begin{table}
    \caption{
        Polynomial coefficients for \teff\ to spectral type relation in equation~(\ref{eq:sptpoly}).
        Valid for $1150 < \teff < 2700$\,K or M6--T4.
    }
    \label{tab:polyorders}
    \begin{tabular}{l|c|c|l}
        a & $6.38\pm1.07$ & $10^{-12}$ & K$^{-4}$ \\
        b & $5.61\pm0.88$ & $10^{-8}$ & K$^{-3}$ \\
        c & $1.83\pm0.27$ & $10^{-4}$ & K$^{-2}$ \\
        d & $2.71\pm0.35$ & $10^{-1}$ & K$^{-1}$ \\
        e & $227\pm17$ &  & K \\
    \end{tabular}
\end{table}

\subsection{Creating a colour ratio} \label{subsec:colourratios}
Following literature definitions of spectral indices in the optical regime\footnote{\label{fn:indices}
Most spectral indices for UCDs are defined in the near infrared rather than the optical, 
see~\citet{2001AJ....121.1710R, 2006ApJ...637.1067B, 2014ApJ...794..143B}, and references therein.
}~\citep{1999ApJ...519..802K, 1999AJ....118.2466M, 2002ApJ...564..466G},
we created a method for measuring a colour ratio (\crall).
This method used directly the \teffespucd\ values in bins of 100\,K\@.
We note here that one spectral type is not equivalent to 100\,K, i.e.\ $\Delta 100\,\mathrm{K} \neq \Delta\ 1 \mathrm{SpT}$.
As for the change in terminology from `spectral index' to `colour ratio', this is because
the internally calibrated \gaia\ RP spectra as shown in Fig.~\ref{fig:gucdsmedians}
are too low resolution to use standard spectral typing indices.
This method created photometric bands centered on the two primary peaks one can see in the
internally calibrated \gaia\ RP spectra (Fig.~\ref{fig:gucdsmedians}).
\citet{2023A&A...674A..33G} discuss the creation of synthetic photometry from \gaia\ XP spectra,
which inspired our method.
Due to the redshifting of these peaks with decreasing effective temperature we define two spectral
\teff-specific narrow bands (with width $50$\,nm), named `\blue' and `\red' respectively,
where the central wavelength shifts with spectral type.
These central wavelengths are the vertical dashed lines shown in Fig.~\ref{fig:gucdsmedians}.
We linearly interpolate between each manually defined central wavelength against \teff\ to
account for the non-rounded \teff\ values.
The total region possibly bound by this relation is 795--995\,nm, i.e.\ the lowest and highest wavelength
within 25\,nm of the central wavelengths.

\begin{figure}
    \includegraphics[width=\linewidth]{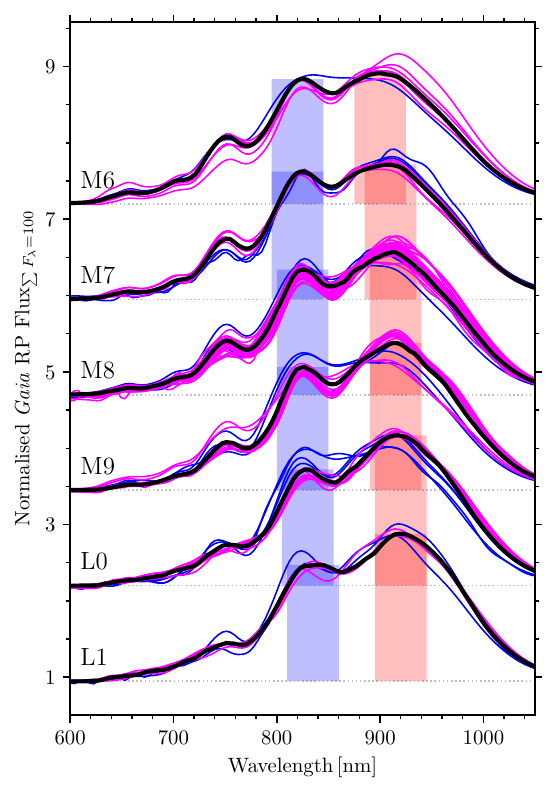}
    \caption{
        Internally calibrated RP spectra of known objects, separated by their literature optical spectral types.
        Magenta spectra are known young objects whilst blue spectra are known subdwarfs.
        Over-plotted in black is the median RP spectra for a given spectral type from known objects in the GUCDS\@.
        The \blue\ and \red\ bands are shown in their respective positions and colours as described in
        Section.~\ref{subsec:colourratios}.
        The normalised spectra were multiplied by a constant value such that the fluxes sum to 100 instead of 1
        and are offset by a set value.
    }
    \label{fig:knownobjects}
\end{figure}

These regions were decided by visually inspecting the known standards, subdwarfs and young objects from
the literature (Fig.~\ref{fig:knownobjects}).
The flux summed in \blue, divided by the flux summed in \red\ can be deemed a `\colour'.
To create \crall\ we had to compare an object's observed \colour\ to an `expected' \colour.

We constructed a median RP normalised spectrum for every $100$\,K bin (using the \gaia\ \teff, \teffespucd).
Then we determined the \colour\ for each median (i.e.\ the `expected' \colour).
We created a linear spline relation between \teff\ and this expected \colour.
Then, for every object, we measure the observed \colour\ and compare it to the expected \colour, extracted
from the linear spline for that object's \teff.
\crall\ is each object's observed \colour\ divided by the expected \colour, rounded to two decimal places.

We sought outliers from \crall\ to define candidate objects.
Values of \crall\ near 1 mean that object is normal.
The median RP spectra of known objects are shown in Fig.~\ref{fig:gucdsmedians}, having been selected from the GUCDS
by each spectral type bin from M5--T6.
We used median RP spectra instead of the known standards in our \crall\ derivation method because of the larger amount
of objects and wider spectral coverage, with the numbers of objects per spectral type
bin shown in Fig.~\ref{fig:gucdshist}.
In our \colour\ region, the median RP spectra per spectral type differ from the known standards
by $|\Delta F| \leqslant 10$\,per cent.
The major caveat for this method is that the \teffespucd\ values were generated from a training set which contained
no outliers.
Hence, it can be expected to be biased.
We may be comparing an observed \colour\ against expectations from an incorrect bin.

\subsubsection{Determining outliers} \label{subsubsec:candidates}
For each object, the outliers were defined as the cases where \crall\ was more than $3\sigma$ from the
average value $\mu$ of all elements of \crall\ ($\mu = 0.98\pm0.05$).
Assuming a Gaussian distribution ($z$) centered at $\mu$, this $\pm3\sigma$ equated to the 0.01\,per cent
and 99.9\,per cent percentiles ($p$) of $z_p$.
In terms of \crall, the 0.01\,per cent percentile, $z_{-3\sigma}$, equals 0.80 whilst the 99.9\,per cent percentile,
$z_{3\sigma}$, equals 1.16.
To summarise, this outlier selection was $z_{-3\sigma} \geq \crall \geq z_{3\sigma}$ or
$0.80 \geq \crall \geq 1.16$ where $p=\pm3\sigma$.
This process went through multiple iterations of different bin sizes, \blue\ and \red\ definitions
(e.g.\ shifting with spectral type and not), numerical methods of creating \crall,
and different \crall\ cut-off points.
We chose the final method parameters such that it only selects the most extreme outliers.
Under this selection criteria, subdwarf candidates were the objects with $\crall \geq z_{3\sigma}$ whilst
young candidates had $\crall \leq z_{-3\sigma}$.

\section{Analysis} \label{sec:dr3analysis}
We discuss here methods of selecting interesting sub-samples of the candidate objects found by the \crall\
in Sect.~\ref{subsubsec:candidates}, although we provide the \crall\ measure for every object.
This analysis section is intended to produce a list of `prime' candidates,
which are the objects passing strict selection criteria.
The aforementioned known standard sample was used to calibrate our \crall\ values,
and ensure we were not selecting `normal' objects.

\begin{figure}
    \includegraphics[width=\linewidth]{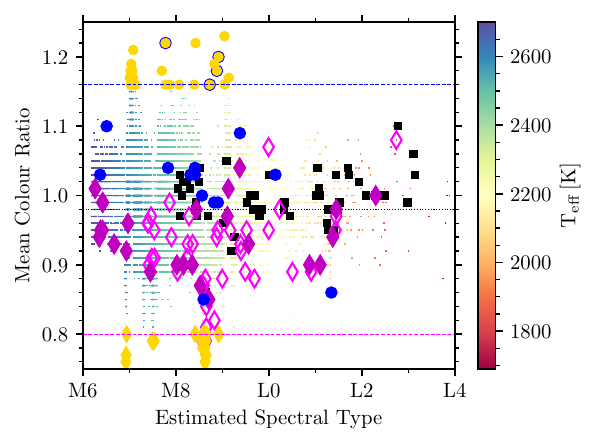}
    \caption{
        Colour ratio (\crall, Sect.~\ref{subsubsec:candidates}) against
        estimated spectral type (Sect.~\ref{subsec:spectyping}).
        We display sources only between M6--L4 (there are no later candidates).
        The full population is shown as small squares using a colour-code reflecting \teff shown on the right-hand axis.
        Standards are displayed as black squares whilst known young objects are magenta diamonds
        (filled if very low gravity, i.e.\ $\delta$ / `vl-g') and known subdwarfs are blue circles.
        Horizontal coloured lines are shown demarcating the selection criteria,
        magenta for $\crall \leq z_{-3\sigma}$ and
        blue for $\crall \geq z_{3\sigma}$.
        A black dotted line is shown at the mean \crall.
        Candidate subdwarfs are yellow circles, candidate young objects are yellow diamonds.
    }
    \label{fig:crselect}
\end{figure}

We defined any object with $\crall \geq z_{3\sigma}$ as
a \crall-candidate subdwarf and anything with $\crall \leq z_{-3\sigma}$ as a \crall-candidate young object.
This selection process is shown in Fig.~\ref{fig:crselect}.

There was an over density of sources around M7--M8, and therefore a less reliable
median RP spectrum, hence the larger \crall\ scatter and artifacts shown in Fig.~\ref{fig:crselect}.
This is due to the artificial upper limit of $\teff < 2700$\,K in \teffespucd.

\begin{figure}
    \includegraphics[width=\linewidth]{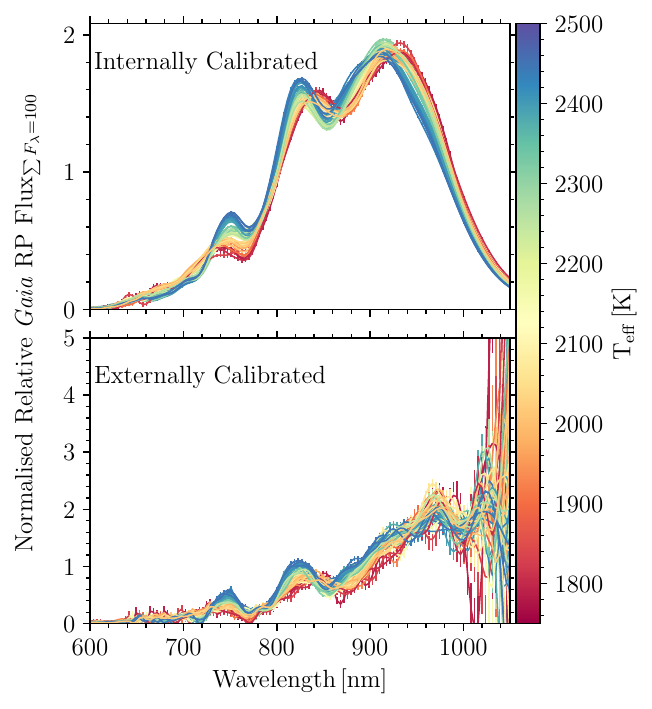}
    \caption{
        Spectral comparison between internally and externally calibrated RP spectra
        of spectral type standards from M7--L4.
        Spectra are coloured by effective temperature.
        Internally calibrated RP spectra of spectral type standards in the upper plot.
        Externally calibrated RP spectra of spectral type standards in the lower plot.
        The normalised spectra were multiplied by a constant value such that the fluxes sum to 100 instead of 1.
    }
    \label{fig:standards_combined}
\end{figure}

Out of \ntotal\ RP spectra, \noutlier\ passed the aforementioned \crall\ cuts.
Following the discussion in section.~3 by~\citet{2023A&A...669A.139S},
we used internally calibrated RP spectra instead of externally calibrated RP spectra.
This is because, as shown by spectral type standards in Fig.~\ref{fig:standards_combined},
the external calibration produces non-physical artifacts for some UCDs~\citep{2021A&A...652A..86C, 2023A&A...674A...3M}.
It was not entirely predictable which objects saw the worst performance in the external calibration;
however, generally the least bright and least observed (\texttt{phot\_rp\_n\_obs}) objects had less reliable spectra.
This is due to the external calibration being derived with sources outside of the UCD
regime~\citep{2012MNRAS.426.1767P}.
\gaia\ observes internally calibrated spectra, not externally calibrated ones.
We base our analysis on a set of spectra that has not undergone an additional calibration stage
which was not optimised for these red and faint sources.
External calibration may introduce systematics upon which we have no control,
in the context of a problem where the signal is very weak.
The internally calibrated RP spectra showed a much cleaner spectral sequence, which was vital for
determining if a given object is `typical' in appearance for a given spectral type, or not.
Both the internal and external calibration spectra were converted from physical wavelengths to
`pseudo-wavelengths' (used by \texttt{gaiaxpy}) via the dispersion function shown in fig.~9
from~\citet{2023A&A...674A...3M} and discussed in section.~3.1 from~\citet{2023A&A...674A...2D}.
This dispersion function is available through \texttt{gaiaxpy} and documented as \href{https://gaiaxpy.readthedocs.io/en/latest/gaiaxpy.calibrator.html\#gaiaxpy.calibrator.external_instrument_model.ExternalInstrumentModel.wl_to_pwl}{\texttt{ExternalInstrumentModel.wl\_to\_pwl}}.
Flux uncertainties were larger in the external calibration, as shown in Fig.~\ref{fig:standards_combined}.
One explanation for this is the known issue in \gdrthree\ that the internal calibration flux uncertainties
are underestimated.
The external calibration did have a larger relative range of fluxes from $F_{\min}$--$F_{\max}$
across our 795--995\,nm region (Sect.~\ref{subsec:colourratios}).
Such a larger relative range would produce improved discernment between neighbouring objects.

\subsection{Photometry checks} \label{subsec:photometry}
\begin{figure*}
    \includegraphics[width=\linewidth]{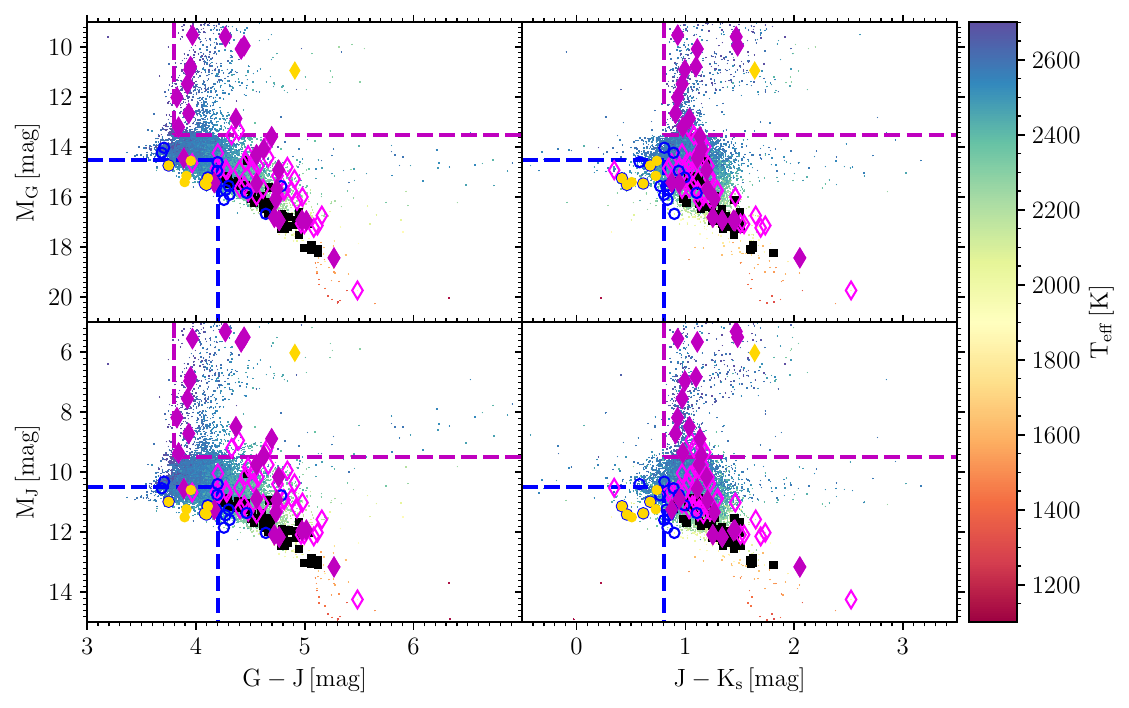}
    \caption{
        Four colour-absolute magnitude diagrams with $M_G$ on the top row, $M_J$ on the bottom row,
        $G-J$ on the left column, and $J-K_s$ on the right column.
        The full RP spectral sample is shown as small squares using a colour-code reflecting \teff,
        as shown in the colour bar.
        Standards are displayed as black squares whilst known young objects are open magenta diamonds
        (filled if very low gravity $\delta$ / `vl-g')
        and known subdwarfs are open blue circles.
        Candidate subdwarfs are yellow circles, candidate young objects are yellow diamonds.
        Dashed lines are shown demarcating the cut-offs for the photometric filtering of the candidate selection.
        Magenta lines are for the young object candidate selection and blue lines are for the
        subdwarf selection.
        These lines represent the cuts in Table~\ref{tab:photocuts}.
    }
    \label{fig:photometry}
\end{figure*}

In the optical regime of \gaia, subdwarfs are known to be typically blue objects
whilst young objects are overluminous and red.
As such, we constructed a CMD to check that candidate objects are in the same
colour-space as known subdwarfs or known young objects.
This is shown in Fig.~\ref{fig:photometry}.
To do this, we created a selection of photometric cuts in Table~\ref{tab:photocuts}.
These are conservative selections on the two categories, aimed at selecting the bluest known 
subdwarfs and brightest known young objects.
We made the selections conservative in order to avoid contaminant
sources, as most contaminants are within the inherent CMD scatter on the
UCD main sequence.

\begin{table}
    \centering
    \caption{
        Photometric cuts to select subdwarfs and young objects.
    }
    \label{tab:photocuts}
    \begin{tabular}{r | r}
        Subdwarf        & Young              \\ \hline
        $M_G > 14.5$    & $M_G < 13.5$       \\
        $G - J < 4.2$   & $G - J > 3.8$      \\
        $M_J > 10.5$    & $M_J < 9.5$        \\
        $J - K_s < 0.8^{\bigstar}$ & $J - K_s \geq 0.8$ \\
    \end{tabular}
    \caption*{$^{\bigstar}$ Slightly more liberal than the $J - K_s < 0.7$ cut by~\citet{2012A&A...542A.105L}.}
\end{table}

There are \nphotoyo\ candidate young objects and \nphotosd\ candidate subdwarfs purely from the photometric cuts
in Table~\ref{tab:photocuts}.
However, only \noutphotyo\ object is both a \crall\ candidate, and a photometric young candidate whilst
\noutphotsd\ objects are both \crall\ candidates, and photometric subdwarf candidates.

\subsection{Kinematics} \label{subsec:kinematics}
We provide a kinematic classification system to indicate thin disc, thick disc, and halo,
based on each object's space motions.
These motions were calculated using the equations from
\texttt{\href{https://github.com/segasai/astrolibpy}{astrolibpy}}, which follows the work by~\citet{1987AJ.....93..864J},
except that U is defined as positive towards the Galactic anti-centre.
We used the Local Standard of Rest (LSR) from~\citet{2011MNRAS.412.1237C} with $\mathrm{U, V, W} = (-8.50, 13.38, 6.49\,\kms)$.
To create UVW velocities, we needed radial velocities to complement the 5-D astrometry from \gdrthree\@.

We cross-matched our sample of \ntotal\ objects with \gaia\ RP spectra 
with SIMBAD~\citep{2000A&AS..143....9W}.
This provided 2187 UCDs with literature radial velocities.
For sources without radial velocities we estimated probability density distributions of the total velocity
by assuming a normal radial velocity distribution.
This distribution was obtained by a maximum likelihood fit to the values available from 
the literature, where $\mu=0.2$\,\kms, $\sigma=52.3$\,\kms.
We sampled 1000 random radial velocities from this normal distribution for each object in our full sample.
Therefore, each object had 1000 different UVW velocities.
This converted into 1000 \vtotal\ values through $\vtotal = \sqrt{U^2 + V^2 + W^2}$.
From each object's range of \vtotal\ values,
we extracted probabilities ($P$) of Galaxy component membership (thin disk, \pthin;
thick disk, \pthick;
halo, \phalo).
This assumes that U, V, W and \vtotal\ are Gaussian distributions propagated
from the normal radial velocity distribution and ignores the impact of metallicity on thick disk/halo discrimination.
To do so, we calculated the \href{https://docs.scipy.org/doc/scipy/reference/generated/
scipy.stats.norm.html#scipy.stats.norm}{survival function}\footnote{\label{fn:survival}
Equivalent to $1 - \text{CDF}$ (Cumulative Distribution Function).}
of each object's total velocity distribution at two critical velocities:
70\,\kms and 180\,\kms~\citep{2010A&A...511L..10N}.
These are checked in descending order:
$\phalo = P(\vtotal > 180\,\text{\kms})$,
$\pthick = \max \{0,\ P(\vtotal > 70\,\text{\kms}) - \phalo\}$,
$\pthin = \max \{0,\ 1. - \pthick - \phalo\}$.
We then select the Galaxy component for each object as whichever probability is highest.

\begin{figure}
    \includegraphics[width=\linewidth]{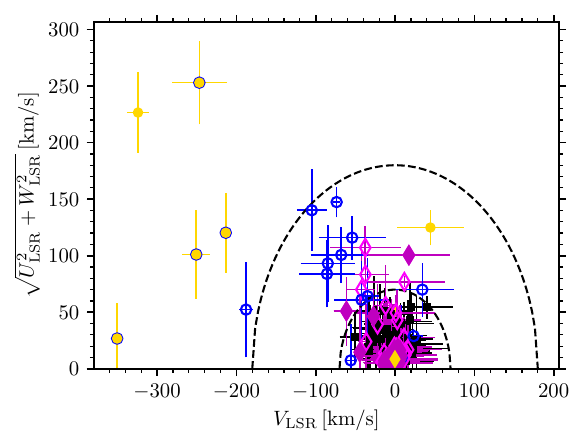}
    \caption{
        Toomre diagram~\citep{1987AJ.....93...74S}, corrected for the LSR,
        of our prime candidates with thick disk and halo selection lines shown
        at $\vtotal > 70\,\kms \text{and}\ \vtotal > 180$\,\kms respectively.
        Standards are displayed as black squares whilst known young objects are open magenta diamonds
        (filled if very low gravity $\delta$ / `vl-g')
        and known subdwarfs are open blue circles.
        Candidate subdwarfs are yellow circles, candidate young objects are yellow diamonds.
        Error-bars in matching colours are also shown.
    }
    \label{fig:dr3toomre}
\end{figure}

Of our candidates, subdwarf candidates were those objects in the halo (\nhalocand)
or thick disk (\nthickcand);
whilst we required young objects to be in the thin disk (although some known young objects can be in the thick disk).
Nevertheless, for young candidates, \nyocand\ object passed all of the respective \crall,
photometric and kinematic cuts.
For the subdwarf candidates, \nsdcand\ objects passed all of the respective \crall, photometric and kinematic cuts.
These \nprimecand\ objects are our prime candidates.
We present the surviving candidates on the Toomre diagram in Fig.~\ref{fig:dr3toomre},
using the mean (of the 1000 total) UVW velocities with propagated uncertainties shown.

\section{Results} \label{sec:dr3results}
\begin{figure*}
    \includegraphics[width=\linewidth]{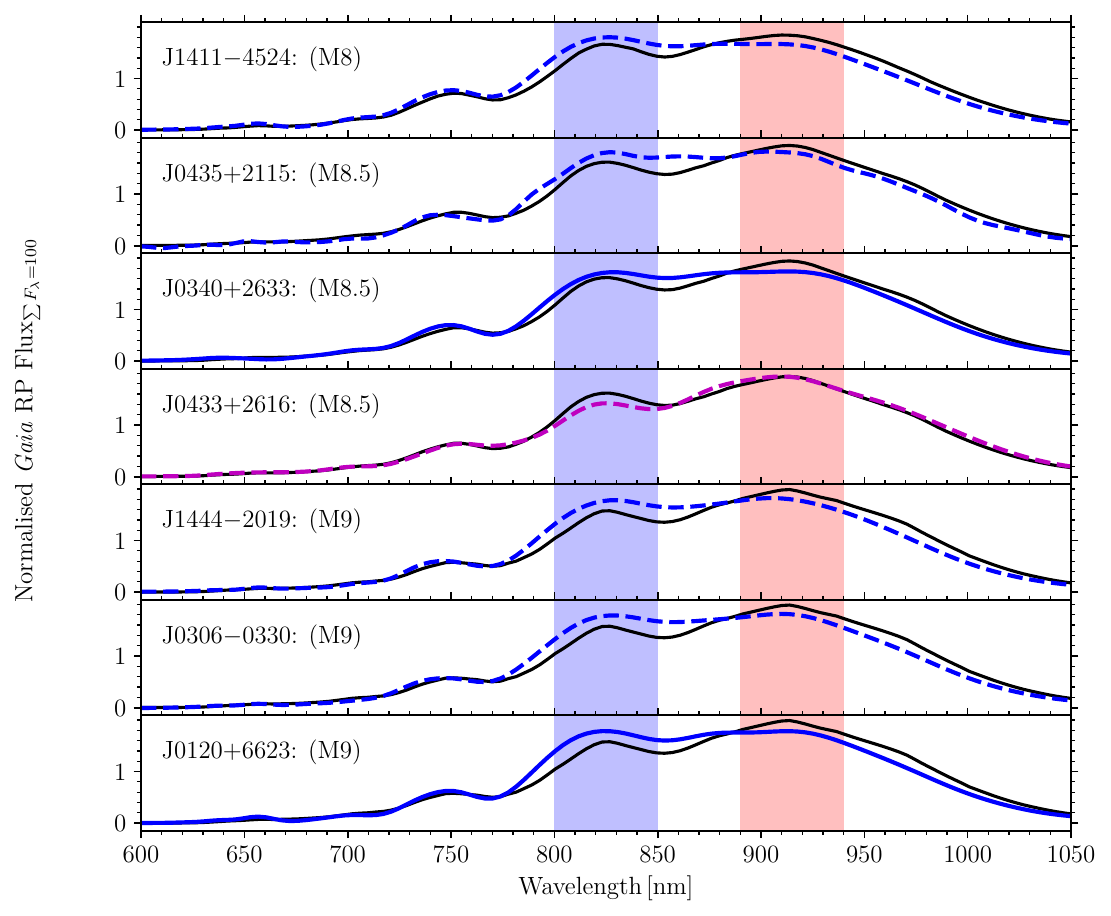}
    \caption{
        Internally calibrated RP spectra of our \nprimecand\ prime candidates
        with estimated spectral type, rounded to 0.5, indicated.
        Any objects with dashed lines are already known to the literature.
        Blue lines are subdwarfs whilst magenta lines are young objects.
        Over-plotted in black is the median RP spectra for the given spectral type from known objects in the GUCDS\@.
        Subdwarfs are typically overluminous in \blue\ and underluminous in \red\ (the \blue\ and
        \red\ bands shown as shaded regions, as described in Sect.~\ref{subsec:colourratios})
        with the inverse true for young objects.
        The normalised spectra were multiplied by a constant value such that the fluxes sum to 100 instead of 1.
    }
    \label{fig:candidatespectra}
\end{figure*}

\begin{figure*}
    \includegraphics[width=\linewidth]{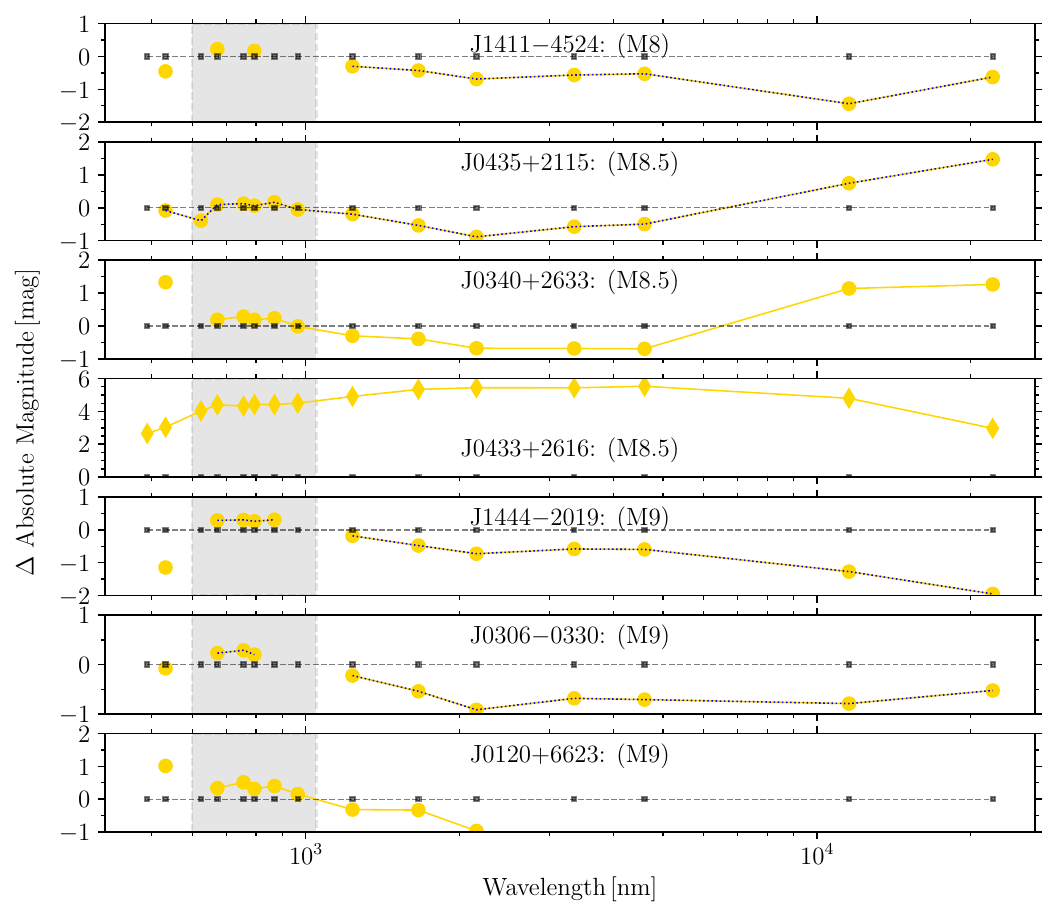}
    \caption{
        The $\Delta$ SEDs for our \nprimecand\ prime candidates in yellow with estimated spectral type 
        (rounded to 0.5) indicated,
        as compared with the mean absolute magnitudes for the given spectral type
        from the GUCDS\@.
        Positive values indicate over-brightness and negative values under-brightness.
        Blue dotted lines are shown on the objects already known to be subdwarfs in the literature.
        Over-plotted in dark grey at zero are the wavelengths covered.
        A grey shading is shown in the region covered by \gaia\ RP spectra.
        The photometry shown is from Pan-STARRS, \gaia, 2MASS and AllWISE\@;
        converted into an absolute magnitude using the \gdrthree\ parallax.
        The wavelengths plotted correspond to the mean wavelengths ($\overline{\lambda}$) of each photometric band
        ($g,\ G_{\text{BP}},\ r,\ G,\ i,\ G_{\text{RP}},\ z,\ y,\ J
        ,\ H,\ K_s,\ W1,\ W2,\ W3,\ W4$, in increasing $\overline{\lambda}$ order), as extracted from
        \href{http://svo2.cab.inta-csic.es/theory/fps/index.php}{VOSA}~\citep{2008A&A...492..277B}.
    }
    \label{fig:sed}
\end{figure*}

We present the \gaia\ RP spectra of the final, \nprimecand\ prime candidates, having survived all \crall,
photometric and kinematic cuts in Fig.~\ref{fig:candidatespectra}
with their astrometry, spectral type and \teff\ shown in Table~\ref{tab:primecandidates}.
We also show the stellar energy distribution (SED) difference from a normal SED of the same spectral type,
for each object in Fig.~\ref{fig:sed}.

\begin{table*}
\centering
\caption{
\label{tab:primecandidates}
Unsorted list of candidate subdwarfs and young objects.
Astrometry is from \gdrthree\ and the \teff\ values are those produced by the \espucd\ Apsis module and 
published as part of the Data Release.
}
\begin{tabular}{c cc c ll c}
\hline \hline
    \gdrthree  & $\alpha$ & $\delta$ & $\varpi$ & Object   & Spectral   & Teff \\
     Source ID & [hms]     & [dms]    & [mas]     & Name     & Type       & [K] \\
\hline
    6281432246412503424 & 14 44 17 & -20 19 56.9 & $58.1\pm0.1$ & SSSPM J1444$-$2019$\hyperlink{candidaterefs}{^{1}}$ & sdM9$\hyperlink{candidaterefs}{^{2}}$ & $2352\pm10$\\
    6096164227899898880 & 14 11 42 & -45 24 20.1 & $19.1\pm0.2$ & 2MASS J14114474$-$4524153$\hyperlink{candidaterefs}{^{3}}$ & sdM9$\hyperlink{candidaterefs}{^{4}}$ & $2487\pm47$\\
    144711230753602048 & 4 35 36 & +21 15 03.6 & $16.7\pm0.6$ & 2MASS J04353511$+$2115201$\hyperlink{candidaterefs}{^{3}}$ & sdL0$\hyperlink{candidaterefs}{^{5}}$ & $2371\pm74$\\
    5183457632811832960 & 3 06 02 & -3 31 06.1 & $24.7\pm0.3$ & 2MASS J03060140$-$0330438$\hyperlink{candidaterefs}{^{3}}$ & sdL0$\hyperlink{candidaterefs}{^{5}}$ & $2348\pm55$\\
    70974545020346240 & 3 40 58 & +26 33 40.8 & $10.6\pm0.7$ & 2MASS J03405673$+$2633447$\hyperlink{candidaterefs}{^{6}}$ & sdM8.5$\hyperlink{candidaterefs}{^{7}}$ & $2411\pm111$\\
    525463551877051136 & 1 20 44 & +66 23 59.0 & $12.1\pm0.4$ & 2MASS J01204397$+$6623543$\hyperlink{candidaterefs}{^{6}}$ & sdM9$\hyperlink{candidaterefs}{^{7}}$ & $2359\pm106$\\
    151130591952773632 & 4 33 08 & +26 16 06.3 & $6.6\pm0.2$ & [BLH2002] KPNO$-$Tau 14$\hyperlink{candidaterefs}{^{8}}$ & M7.2$\hyperlink{candidaterefs}{^{9}}$ & $2385\pm18$\\
    \hline
\end{tabular}
\caption*{\hypertarget{candidaterefs}{References}: 1.~\cite{2004A&A...428L..25S}, 
2.~\cite{2015AJ....149....5W}, 
3.~\cite{2014ApJ...781....4L}, 
4.~\cite{2016ApJS..224...36K}, 
5.~\cite{2014ApJ...783..122K}, 
6.~\cite{2003yCat.2246....0C}, 
7.~This Work, 
8.~\cite{2003ApJ...590..348L}, 
9.~\cite{2018ApJ...858...41Z}}
\end{table*}

We discuss here each object classified as a prime candidate in this work.
Four candidates were already known subdwarfs and flagged as such in the GUCDS:
\begin{itemize}
    \item \textbf{SSSPM J1444$-$2019} \textit{(J1444$-$2019)}:
    In the literature, this object is an M9~\citep{2014ApJ...794..143B} or an
    sdL0~\citep[in both the optical and near-infrared regime,][]{2016ApJS..224...36K}.
    This work estimated a spectral type of M9, $\crall = 1.18$ and $\phalo = 1$.
    Our spectral type agrees with the literature's modal spectral type and our kinematics combined with it's
    blue nature confirm the subdwarf.

    \item \textbf{2MASS J14114474$-$4524153} \textit{(J1411$-$4524)}:
    J1411$-$4524 is an sdM9~\citep{2016ApJS..224...36K}.
    We found a spectral type of M8, $\crall = 1.22$ and $\phalo = 1$, hence our agreed classification as a subdwarf.

    \item \textbf{2MASS J04353511$+$2115201} \textit{(J0435$+$2115)}:
    An sdL0 (optical) object~\citep{2014ApJ...783..122K}, confirmed by~\citet{2016ApJS..224...36K}
    with a similar sdM9 from~\citet{2014ApJ...787..126L}
    \footnote{\label{fn:luhmanfnote}
    There appears to be some confusion in the literature bibliography codes (bibcodes)
    about the origin of this spectral type.
    There are three very similar bibcodes:
    ~\citet[2014ApJ...787..126L -- `Characterization of High Proper
    Motion Objects from the Wide-field Infrared Survey Explorer'][]{2014ApJ...787..126L};
    ~\citet[2014ApJ...786L..18L -- `Discovery of a ${\sim}$250\,K Brown Dwarf at 2\,pc
    from the Sun'][]{2014ApJ...786L..18L};
    ~\citet[2014ApJ...781....4L -- `A Search for a Distant Companion to the Sun with
    the Wide-field Infrared Survey Explorer'][]{2014ApJ...781....4L};
    the correct reference is~\citet{2014ApJ...787..126L}.}.
    The spectral type from this work is M8.5, mostly in agreement with the literature,
    with $\crall = 1.16$ and $\phalo = 1.0$.
    We concur with the subdwarf classification.

    \item \textbf{2MASS J03060140$-$0330438} \textit{(J0306$-$0330)}:
    Similarly, an sdL0 (optical) object~\citep{2014ApJ...783..122K} with an sdM9 sub-type
    from~\citet{2014ApJ...787..126L}\textsuperscript{\ref{fn:luhmanfnote}}.
    This work estimated a spectral type of M9.
    $\crall = 1.20$ and $\phalo = 1.0$, the high \crall\ value indicates this object is a likely subdwarf.
\end{itemize}

Two new subdwarf candidates were also found:
\begin{itemize}
    \item \textbf{2MASS J03405673$+$2633447} \textit{(J0340$+$2633)}:
    Not known to SIMBAD (besides an entry for \gdrthree\ and 2MASS) or the GUCDS\footnote{\label{fn:gucdsinclusion}
    This isn't unexpected, as the GUCDS is only intended to be complete for L dwarfs.}.
    We found a spectral type of M8.5, $\crall = 1.16$ and $\phalo = 1.0$.
    The \crall\ value is on the borderline of the cut-off, however, this is still significant, especially considering
    that it has the fastest $\vtan$ in the sample at 407.3\,\kms.
    It shows a non detection in PS1 $g$ \& $r$ and is generally underluminous in the NIR (Fig.~\ref{fig:sed})
    but overluminous in the two reddest bands of AllWISE, a similar pattern to J0435$+$2115
    (the known subdwarf of the same estimated spectral type).
    The missing detection in PS1 is due to the cross-matching, when visually inspected there is a highly red object
    visible within ${\approx}2$\,arcseconds.
    J0340$+$2633 is even more blue in Fig.~\ref{fig:photometry} than most of our known subdwarfs, as would be
    expected for an extreme object.

    \item \textbf{2MASS J01204397$+$6623543} \textit{(J0120$+$6623)}:
    Likewise, this object has a lack of information in the literature.
    This work estimated a spectral type of M9, with $\crall = 1.19$ and $\pthick = 1.0$.
    The very high \crall\ value also indicates this object is also non-standard for an M9\@.
    It also shows a non detection in PS1 $g$ \& $r$ but additionally no match in AllWISE\@.
    This is again due to the cross-matching uncertainties as there is a clear red object in PS1 when visually inspected.
    It appears in the AllWISE images that the object is hidden by two neighbouring bright stars.
    However, it is \textit{tending} towards being underluminous in the NIR (Fig.~\ref{fig:sed}),
    as would be expected from the two known subdwarfs of the same
    estimated spectral type (J1444$-$2019 and J0306$-$0330).
    As with J0340$+$2633, J0120$+$6623 is notably more blue than other subdwarfs known to the literature in 
    Fig.~\ref{fig:photometry}.
    This is therefore classed as a new subdwarf.
\end{itemize}

Additionally, we found one young object candidate, already known to the literature:
\begin{itemize}
    \item \textbf{[BLH2002] KPNO$-$Tau~14} \textit{(J0433$+$2616)}:
    This object is not in the GUCDS\textsuperscript{\ref{fn:gucdsinclusion}}
    but is an M7.2~\citep{2018ApJ...858...41Z} in SIMBAD and classed as M6Ve by~\citet{2003ApJ...590..348L}\@.
    \citet{2019AJ....157..196K} gives this object a radial velocity of $17.07\pm0.37$,
    which combined with the $\vtan$ of $13.84$\,\kms, suggests it is strongly within the thin disk.
    It has also been repeatedly shown to be a member of the Taurus star forming 
    complex~\citep{2006ApJ...647.1180L, 2007ApJ...662..413K, 2010ApJS..186..111L, 2010ApJS..186..259R,
    2018AJ....156..271L, 2020AJ....159..273R} and generally within the
    Taurus-Auriga ecosystem~\citep{2017ApJ...838..150K}.
    It is a young stellar object (YSO) with an age (from membership of Taurus) of 1--2\,Myr~\citep{2018ApJ...856...23G}.
    Our spectral type is M8.5, within $2\sigma$ of the literature values, which is most likely due to the \teff\ scatter
    in that spectral type bin (see Fig.~\ref{fig:sptconversion}), in addition to the fact that YSOs are highly variable.
    The $\crall = 0.83$ and $\pthin = 0.8$.
    Figure~\ref{fig:sed} shows this object is significantly overluminous for it's
    spectral type, again typical of a YSO\@.
\end{itemize}

\section{Discussion} \label{sec:dr3discussion}
This work has produced a list of \noutlier\ objects, which have \gaia\ RP spectral differences greater than $3\sigma$
from median RP spectra, derived using the GUCDS and a new colour ratio (\crall) specific to
internally calibrated \gaia\ RP spectra.
We finally produced a list of \nprimecand\ prime candidates,
which have passed highly restrictive photometric and kinematic selections,
aimed at recovering the most extreme objects in the sample.

Whilst we could have used a more liberal set of cuts, the intention in this work was to produce the most confident candidates.
Additionally, part of the publication criteria (see Sect.~\ref{sec:dr3method}) for \gaia\ RP UCD
spectra was that the RP spectra had the highest quality flags (\flagsespucd\ 0--1).
This meant objects with higher Euclidean distances from
BT-Settl~\citep{2011ASPC..448...91A} models (simulated through the \gaia\ RP transmission function) are not included.
In other words, the most extreme objects we seek to classify were precluded from inclusion in \gdrthree
\footnote{\label{fn:qflagespucd}
However, the quality flag selections performed by \espucd\ were very 
sensible, see discussion by~\citet{2023A&A...674A..26C} and~\citet{2023A&A...669A.139S},
as there were many potential contaminants and highly noisy spectra in the lowest quality flag (2).
}\@.

Several other biases exist, such as the artificial cut of $\teff < 2700$\,K from \teffespucd.
This caused the over density seen at the M7--M8.
The lack of outliers in the empirical training set in \gdrthree\ also caused a bias in the creation of expected \colour.
Also, the sample of known young objects and known subdwarfs in the GUCDS includes many objects, which appear not
considerably different from a normal object when visually observed at a resolution
as low as \gaia\ RP, see Fig.~\ref{fig:knownobjects}.
This can be evidenced by Fig.~\ref{fig:crselect}, where there is little scatter in \crall\ in spectral sub-types
beyond L0.
These objects are as equally interesting as extreme outliers, but require higher resolution optical and NIR spectroscopy 
to observe directly the features relating to surface gravity and metallicity.
Many of these objects did not pass the \crall\ selection, photometric and kinematic cuts, or both.
These reasons combined with the rarity of extreme UCDs are the cause of there being so few prime candidates in our final list.
However, the detection of the known extreme UCDs shown here is a highly promising baseline for future analysis.
The additional detection of two unknown subdwarf candidates is demonstrative of the fact that existing datasets, like
\gdrthree, contain many interesting objects, still to be discovered.
This future work could include more advanced selection techniques such as machine learning, more liberal selection
criteria and the increased breadth and depth of planned \gaia\ data releases.

\subsection*{Data availability}
The data underlying this article will be shared on reasonable request to the corresponding author.
It will additionally be available through \href{https://vizier.cds.unistra.fr/viz-bin/VizieR}{CDS VizieR}.

\section*{Acknowledgements}
RLS and WJC were been supported by a STSM grant from COST Action CA18104: MW-Gaia.
The authors would like to thank Jos\'e Caballero at ESAC for his much appreciated advice.
WJC was funded by a University of Hertfordshire studentship and both WJC and HRAJ were supported by STFC
grant ST/R000905/1 at the University of Hertfordshire.
We would like to thank the anonymous reviewer for their timely and useful advice,
which has much improved this technique and manuscript.

This publication makes use of reduction and data products from the Centre de Données astronomiques de Strasbourg
(SIMBAD, \url{cdsweb.u-strasbg.fr});
the ESA Gaia mission (\url{http://sci.esa.int/gaia}) funded by national institutions participating in the
Gaia Multilateral Agreement and in particular the support of ASI under contract I/058/10/0 (Gaia Mission -
The Italian Participation to DPAC);
the Panoramic Survey Telescope and Rapid Response System (Pan-STARRS, \url{panstarrs.stsci.edu});
the Sloan Digital Sky Survey (SDSS, \url{www.sdss.org});
the Two Micron All Sky Survey (2MASS, \url{www.ipac.caltech.edu/2mass}) and the
Wide-field Infrared Survey Explorer (WISE, \url{wise.ssl.berkeley.edu}).

We acknowledge the relevant open source packages used in our
\texttt{python}~\citep{python} codes:
\texttt{astropy}~\citep{astropy:2013, astropy:2018},
\texttt{matplotlib}~\citep{Hunter:2007},
\texttt{numpy}~\citep{harris2020array},
\texttt{pandas}~\citep{mckinney-proc-scipy-2010, reback2020pandas},
\texttt{scipy}~\citep{2020SciPy-NMeth},
\texttt{sympy}~\citep{10.7717/peerj-cs.103},
\texttt{tqdm}~\citep{casper_da_costa_luis_2021_5517697}.

\bibliographystyle{mnras}
\bibliography{references}

\begin{thebibliography}{}
\makeatletter
\relax
\def\mn@urlcharsother{\let\do\@makeother \do\$\do\&\do\#\do\^\do\_\do\%\do\~}
\def\mn@doi{\begingroup\mn@urlcharsother \@ifnextchar [ {\mn@doi@}
  {\mn@doi@[]}}
\def\mn@doi@[#1]#2{\def\@tempa{#1}\ifx\@tempa\@empty \href
  {http://dx.doi.org/#2} {doi:#2}\else \href {http://dx.doi.org/#2} {#1}\fi
  \endgroup}
\def\mn@eprint#1#2{\mn@eprint@#1:#2::\@nil}
\def\mn@eprint@arXiv#1{\href {http://arxiv.org/abs/#1} {{\tt arXiv:#1}}}
\def\mn@eprint@dblp#1{\href {http://dblp.uni-trier.de/rec/bibtex/#1.xml}
  {dblp:#1}}
\def\mn@eprint@#1:#2:#3:#4\@nil{\def\@tempa {#1}\def\@tempb {#2}\def\@tempc
  {#3}\ifx \@tempc \@empty \let \@tempc \@tempb \let \@tempb \@tempa \fi \ifx
  \@tempb \@empty \def\@tempb {arXiv}\fi \@ifundefined
  {mn@eprint@\@tempb}{\@tempb:\@tempc}{\expandafter \expandafter \csname
  mn@eprint@\@tempb\endcsname \expandafter{\@tempc}}}

\bibitem[\protect\citeauthoryear{{Aganze} et~al.,}{{Aganze}
  et~al.}{2016}]{2016AJ....151...46A}
{Aganze} C.,  et~al., 2016, \mn@doi [\aj] {10.3847/0004-6256/151/2/46}, \href
  {https://ui.adsabs.harvard.edu/abs/2016AJ....151...46A} {151, 46}

\bibitem[\protect\citeauthoryear{{Allard}, {Homeier}  \& {Freytag}}{{Allard}
  et~al.}{2011}]{2011ASPC..448...91A}
{Allard} F.,  {Homeier} D.,   {Freytag} B.,  2011, in {Johns-Krull} C.,
  {Browning} M.~K.,   {West} A.~A.,  eds,  Astronomical Society of the Pacific
  Conference Series Vol. 448, 16th Cambridge Workshop on Cool Stars, Stellar
  Systems, and the Sun. p.~91 (\mn@eprint {arXiv} {1011.5405})

\bibitem[\protect\citeauthoryear{{Allers} \& {Liu}}{{Allers} \&
  {Liu}}{2013}]{2013ApJ...772...79A}
{Allers} K.~N.,  {Liu} M.~C.,  2013, \mn@doi [\apj]
  {10.1088/0004-637X/772/2/79}, \href
  {https://ui.adsabs.harvard.edu/abs/2013ApJ...772...79A} {772, 79}

\bibitem[\protect\citeauthoryear{{Anders}, {Khalatyan}, {Queiroz}, {Nepal}  \&
  {Chiappini}}{{Anders} et~al.}{2023}]{2023arXiv230206995A}
{Anders} F.,  {Khalatyan} A.,  {Queiroz} A.~B.~A.,  {Nepal} S.,   {Chiappini}
  C.,  2023, \mn@doi [arXiv e-prints] {10.48550/arXiv.2302.06995}, \href
  {https://ui.adsabs.harvard.edu/abs/2023arXiv230206995A} {p. arXiv:2302.06995}

\bibitem[\protect\citeauthoryear{{Andrae}, {Rix}  \& {Chandra}}{{Andrae}
  et~al.}{2023}]{2023ApJS..267....8A}
{Andrae} R.,  {Rix} H.-W.,   {Chandra} V.,  2023, \mn@doi [\apjs]
  {10.3847/1538-4365/acd53e}, \href
  {https://ui.adsabs.harvard.edu/abs/2023ApJS..267....8A} {267, 8}

\bibitem[\protect\citeauthoryear{{Andrei} et~al.,}{{Andrei}
  et~al.}{2011}]{2011AJ....141...54A}
{Andrei} A.~H.,  et~al., 2011, \mn@doi [\aj] {10.1088/0004-6256/141/2/54},
  \href {https://ui.adsabs.harvard.edu/abs/2011AJ....141...54A} {141, 54}

\bibitem[\protect\citeauthoryear{{Ardila}, {Mart{\'\i}n}  \& {Basri}}{{Ardila}
  et~al.}{2000}]{2000AJ....120..479A}
{Ardila} D.,  {Mart{\'\i}n} E.,   {Basri} G.,  2000, \mn@doi [\aj]
  {10.1086/301443}, \href
  {https://ui.adsabs.harvard.edu/abs/2000AJ....120..479A} {120, 479}

\bibitem[\protect\citeauthoryear{{Astropy Collaboration} et~al.,}{{Astropy
  Collaboration} et~al.}{2013}]{astropy:2013}
{Astropy Collaboration} et~al., 2013, \mn@doi [\aap]
  {10.1051/0004-6361/201322068}, \href
  {http://adsabs.harvard.edu/abs/2013A%26A...558A..33A} {558, A33}

\bibitem[\protect\citeauthoryear{{Astropy Collaboration} et~al.,}{{Astropy
  Collaboration} et~al.}{2018}]{astropy:2018}
{Astropy Collaboration} et~al., 2018, \mn@doi [\aj] {10.3847/1538-3881/aabc4f},
  \href {https://ui.adsabs.harvard.edu/abs/2018AJ....156..123A} {156, 123}

\bibitem[\protect\citeauthoryear{{Bardalez Gagliuffi} et~al.,}{{Bardalez
  Gagliuffi} et~al.}{2014}]{2014ApJ...794..143B}
{Bardalez Gagliuffi} D.~C.,  et~al., 2014, \mn@doi [\apj]
  {10.1088/0004-637X/794/2/143}, \href
  {https://ui.adsabs.harvard.edu/abs/2014ApJ...794..143B} {794, 143}

\bibitem[\protect\citeauthoryear{{Bayo}, {Rodrigo}, {Barrado Y Navascu{\'e}s},
  {Solano}, {Guti{\'e}rrez}, {Morales-Calder{\'o}n}  \& {Allard}}{{Bayo}
  et~al.}{2008}]{2008A&A...492..277B}
{Bayo} A.,  {Rodrigo} C.,  {Barrado Y Navascu{\'e}s} D.,  {Solano} E.,
  {Guti{\'e}rrez} R.,  {Morales-Calder{\'o}n} M.,   {Allard} F.,  2008, \mn@doi
  [\aap] {10.1051/0004-6361:200810395}, \href
  {https://ui.adsabs.harvard.edu/abs/2008A&A...492..277B} {492, 277}

\bibitem[\protect\citeauthoryear{{Bouy}, {Brandner}, {Mart{\'\i}n}, {Delfosse},
  {Allard}  \& {Basri}}{{Bouy} et~al.}{2003}]{2003AJ....126.1526B}
{Bouy} H.,  {Brandner} W.,  {Mart{\'\i}n} E.~L.,  {Delfosse} X.,  {Allard} F.,
   {Basri} G.,  2003, \mn@doi [\aj] {10.1086/377343}, \href
  {https://ui.adsabs.harvard.edu/abs/2003AJ....126.1526B} {126, 1526}

\bibitem[\protect\citeauthoryear{{Burgasser}}{{Burgasser}}{2004}]{2004ApJ...614L..73B}
{Burgasser} A.~J.,  2004, \mn@doi [\apjl] {10.1086/425418}, \href
  {https://ui.adsabs.harvard.edu/abs/2004ApJ...614L..73B} {614, L73}

\bibitem[\protect\citeauthoryear{{Burgasser} et~al.,}{{Burgasser}
  et~al.}{2002}]{2002ApJ...564..421B}
{Burgasser} A.~J.,  et~al., 2002, \mn@doi [\apj] {10.1086/324033}, \href
  {https://ui.adsabs.harvard.edu/abs/2002ApJ...564..421B} {564, 421}

\bibitem[\protect\citeauthoryear{{Burgasser}, {McElwain}, {Kirkpatrick},
  {Cruz}, {Tinney}  \& {Reid}}{{Burgasser} et~al.}{2004}]{2004AJ....127.2856B}
{Burgasser} A.~J.,  {McElwain} M.~W.,  {Kirkpatrick} J.~D.,  {Cruz} K.~L.,
  {Tinney} C.~G.,   {Reid} I.~N.,  2004, \mn@doi [\aj] {10.1086/383549}, \href
  {https://ui.adsabs.harvard.edu/abs/2004AJ....127.2856B} {127, 2856}

\bibitem[\protect\citeauthoryear{{Burgasser}, {Geballe}, {Leggett},
  {Kirkpatrick}  \& {Golimowski}}{{Burgasser}
  et~al.}{2006}]{2006ApJ...637.1067B}
{Burgasser} A.~J.,  {Geballe} T.~R.,  {Leggett} S.~K.,  {Kirkpatrick} J.~D.,
  {Golimowski} D.~A.,  2006, \mn@doi [\apj] {10.1086/498563}, \href
  {https://ui.adsabs.harvard.edu/abs/2006ApJ...637.1067B} {637, 1067}

\bibitem[\protect\citeauthoryear{{Carrasco} et~al.,}{{Carrasco}
  et~al.}{2021}]{2021A&A...652A..86C}
{Carrasco} J.~M.,  et~al., 2021, \mn@doi [\aap] {10.1051/0004-6361/202141249},
  \href {https://ui.adsabs.harvard.edu/abs/2021A&A...652A..86C} {652, A86}

\bibitem[\protect\citeauthoryear{{Chambers} et~al.,}{{Chambers}
  et~al.}{2016}]{2016arXiv161205560C}
{Chambers} K.~C.,  et~al., 2016, arXiv e-prints, \href
  {https://ui.adsabs.harvard.edu/abs/2016arXiv161205560C} {p. arXiv:1612.05560}

\bibitem[\protect\citeauthoryear{{Cieza} \& {Baliber}}{{Cieza} \&
  {Baliber}}{2006}]{2006ApJ...649..862C}
{Cieza} L.,  {Baliber} N.,  2006, \mn@doi [\apj] {10.1086/506342}, \href
  {https://ui.adsabs.harvard.edu/abs/2006ApJ...649..862C} {649, 862}

\bibitem[\protect\citeauthoryear{{Co{\c{s}}kuno{\v{g}}lu}
  et~al.,}{{Co{\c{s}}kuno{\v{g}}lu} et~al.}{2011}]{2011MNRAS.412.1237C}
{Co{\c{s}}kuno{\v{g}}lu} B.,  et~al., 2011, \mn@doi [\mnras]
  {10.1111/j.1365-2966.2010.17983.x}, \href
  {https://ui.adsabs.harvard.edu/abs/2011MNRAS.412.1237C} {412, 1237}

\bibitem[\protect\citeauthoryear{Cooper}{Cooper}{2022}]{cooper_w_j_2022_6653446}
Cooper W.~J.,  2022, gaiaxpy-batch, \mn@doi{10.5281/zenodo.6653446}, \url
  {https://doi.org/10.5281/zenodo.6653446}

\bibitem[\protect\citeauthoryear{{Creevey} et~al.,}{{Creevey}
  et~al.}{2023}]{2023A&A...674A..26C}
{Creevey} O.~L.,  et~al., 2023, \mn@doi [\aap] {10.1051/0004-6361/202243688},
  \href {https://ui.adsabs.harvard.edu/abs/2023A&A...674A..26C} {674, A26}

\bibitem[\protect\citeauthoryear{{Cruz}, {Reid}, {Liebert}, {Kirkpatrick}  \&
  {Lowrance}}{{Cruz} et~al.}{2003}]{2003AJ....126.2421C}
{Cruz} K.~L.,  {Reid} I.~N.,  {Liebert} J.,  {Kirkpatrick} J.~D.,   {Lowrance}
  P.~J.,  2003, \mn@doi [\aj] {10.1086/378607}, \href
  {https://ui.adsabs.harvard.edu/abs/2003AJ....126.2421C} {126, 2421}

\bibitem[\protect\citeauthoryear{{Cruz} et~al.,}{{Cruz}
  et~al.}{2007}]{2007AJ....133..439C}
{Cruz} K.~L.,  et~al., 2007, \mn@doi [\aj] {10.1086/510132}, \href
  {https://ui.adsabs.harvard.edu/abs/2007AJ....133..439C} {133, 439}

\bibitem[\protect\citeauthoryear{{Cruz}, {Kirkpatrick}  \& {Burgasser}}{{Cruz}
  et~al.}{2009}]{2009AJ....137.3345C}
{Cruz} K.~L.,  {Kirkpatrick} J.~D.,   {Burgasser} A.~J.,  2009, \mn@doi [\aj]
  {10.1088/0004-6256/137/2/3345}, \href
  {https://ui.adsabs.harvard.edu/abs/2009AJ....137.3345C} {137, 3345}

\bibitem[\protect\citeauthoryear{{Cruz}, {Galindo}, {Faherty}, {Riedel}  \&
  {BDNYC}}{{Cruz} et~al.}{2016}]{2016AAS...22714503C}
{Cruz} K.~L.,  {Galindo} C.,  {Faherty} J.~K.,  {Riedel} A.~R.,   {BDNYC} 2016,
  in American Astronomical Society Meeting Abstracts \#227. p. 145.03

\bibitem[\protect\citeauthoryear{{Culpan}, {Geier}, {Reindl}, {Pelisoli},
  {Gentile Fusillo}  \& {Vorontseva}}{{Culpan}
  et~al.}{2022}]{2022A&A...662A..40C}
{Culpan} R.,  {Geier} S.,  {Reindl} N.,  {Pelisoli} I.,  {Gentile Fusillo} N.,
   {Vorontseva} A.,  2022, \mn@doi [\aap] {10.1051/0004-6361/202243337}, \href
  {https://ui.adsabs.harvard.edu/abs/2022A&A...662A..40C} {662, A40}

\bibitem[\protect\citeauthoryear{{Cushing} et~al.,}{{Cushing}
  et~al.}{2011}]{2011ApJ...743...50C}
{Cushing} M.~C.,  et~al., 2011, \mn@doi [\apj] {10.1088/0004-637X/743/1/50},
  \href {https://ui.adsabs.harvard.edu/abs/2011ApJ...743...50C} {743, 50}

\bibitem[\protect\citeauthoryear{{Cutri} et~al.,}{{Cutri}
  et~al.}{2003}]{2003yCat.2246....0C}
{Cutri} R.~M.,  et~al., 2003, VizieR Online Data Catalog, \href
  {https://ui.adsabs.harvard.edu/abs/2003yCat.2246....0C} {p. II/246}

\bibitem[\protect\citeauthoryear{{Cutri} et~al.,}{{Cutri}
  et~al.}{2013}]{2014yCat.2328....0C}
{Cutri} R.~M.,  et~al., 2013, VizieR Online Data Catalog, \href
  {https://ui.adsabs.harvard.edu/abs/2014yCat.2328....0C} {p. II/328}

\bibitem[\protect\citeauthoryear{{De Angeli} et~al.,}{{De Angeli}
  et~al.}{2023}]{2023A&A...674A...2D}
{De Angeli} F.,  et~al., 2023, \mn@doi [\aap] {10.1051/0004-6361/202243680},
  \href {https://ui.adsabs.harvard.edu/abs/2023A&A...674A...2D} {674, A2}

\bibitem[\protect\citeauthoryear{{Deacon} \& {Hambly}}{{Deacon} \&
  {Hambly}}{2007}]{2007A&A...468..163D}
{Deacon} N.~R.,  {Hambly} N.~C.,  2007, \mn@doi [\aap]
  {10.1051/0004-6361:20066844}, \href
  {https://ui.adsabs.harvard.edu/abs/2007A&A...468..163D} {468, 163}

\bibitem[\protect\citeauthoryear{{Dupuy} \& {Liu}}{{Dupuy} \&
  {Liu}}{2012}]{2012ApJS..201...19D}
{Dupuy} T.~J.,  {Liu} M.~C.,  2012, \mn@doi [\apjs]
  {10.1088/0067-0049/201/2/19}, \href
  {https://ui.adsabs.harvard.edu/abs/2012ApJS..201...19D} {201, 19}

\bibitem[\protect\citeauthoryear{{EROS Collaboration} et~al.,}{{EROS
  Collaboration} et~al.}{1999}]{1999A&A...351L...5E}
{EROS Collaboration} et~al., 1999, \aap, \href
  {https://ui.adsabs.harvard.edu/abs/1999A&A...351L...5E} {351, L5}

\bibitem[\protect\citeauthoryear{{Esplin}, {Luhman}  \& {Mamajek}}{{Esplin}
  et~al.}{2014}]{2014ApJ...784..126E}
{Esplin} T.~L.,  {Luhman} K.~L.,   {Mamajek} E.~E.,  2014, \mn@doi [\apj]
  {10.1088/0004-637X/784/2/126}, \href
  {https://ui.adsabs.harvard.edu/abs/2014ApJ...784..126E} {784, 126}

\bibitem[\protect\citeauthoryear{{Faherty} et~al.,}{{Faherty}
  et~al.}{2012}]{2012ApJ...752...56F}
{Faherty} J.~K.,  et~al., 2012, \mn@doi [\apj] {10.1088/0004-637X/752/1/56},
  \href {https://ui.adsabs.harvard.edu/abs/2012ApJ...752...56F} {752, 56}

\bibitem[\protect\citeauthoryear{{Gagn{\'e}} \& {Faherty}}{{Gagn{\'e}} \&
  {Faherty}}{2018}]{2018ApJ...862..138G}
{Gagn{\'e}} J.,  {Faherty} J.~K.,  2018, \mn@doi [\apj]
  {10.3847/1538-4357/aaca2e}, \href
  {https://ui.adsabs.harvard.edu/abs/2018ApJ...862..138G} {862, 138}

\bibitem[\protect\citeauthoryear{{Gagn{\'e}}, {Faherty}, {Cruz},
  {Lafreni{\`e}re}, {Doyon}, {Malo}  \& {Artigau}}{{Gagn{\'e}}
  et~al.}{2014}]{2014ApJ...785L..14G}
{Gagn{\'e}} J.,  {Faherty} J.~K.,  {Cruz} K.,  {Lafreni{\`e}re} D.,  {Doyon}
  R.,  {Malo} L.,   {Artigau} {\'E}.,  2014, \mn@doi [\apjl]
  {10.1088/2041-8205/785/1/L14}, \href
  {https://ui.adsabs.harvard.edu/abs/2014ApJ...785L..14G} {785, L14}

\bibitem[\protect\citeauthoryear{{Gagn{\'e}} et~al.,}{{Gagn{\'e}}
  et~al.}{2015a}]{2015ApJS..219...33G}
{Gagn{\'e}} J.,  et~al., 2015a, \mn@doi [\apjs] {10.1088/0067-0049/219/2/33},
  \href {https://ui.adsabs.harvard.edu/abs/2015ApJS..219...33G} {219, 33}

\bibitem[\protect\citeauthoryear{{Gagn{\'e}}, {Lafreni{\`e}re}, {Doyon}, {Malo}
   \& {Artigau}}{{Gagn{\'e}} et~al.}{2015b}]{2015ApJ...798...73G}
{Gagn{\'e}} J.,  {Lafreni{\`e}re} D.,  {Doyon} R.,  {Malo} L.,   {Artigau}
  {\'E}.,  2015b, \mn@doi [\apj] {10.1088/0004-637X/798/2/73}, \href
  {https://ui.adsabs.harvard.edu/abs/2015ApJ...798...73G} {798, 73}

\bibitem[\protect\citeauthoryear{{Gagn{\'e}} et~al.,}{{Gagn{\'e}}
  et~al.}{2017}]{2017ApJS..228...18G}
{Gagn{\'e}} J.,  et~al., 2017, \mn@doi [\apjs] {10.3847/1538-4365/228/2/18},
  \href {https://ui.adsabs.harvard.edu/abs/2017ApJS..228...18G} {228, 18}

\bibitem[\protect\citeauthoryear{{Gagn{\'e}} et~al.,}{{Gagn{\'e}}
  et~al.}{2018}]{2018ApJ...856...23G}
{Gagn{\'e}} J.,  et~al., 2018, \mn@doi [\apj] {10.3847/1538-4357/aaae09}, \href
  {https://ui.adsabs.harvard.edu/abs/2018ApJ...856...23G} {856, 23}

\bibitem[\protect\citeauthoryear{{Gaia Collaboration} et~al.,}{{Gaia
  Collaboration} et~al.}{2016}]{2016A&A...595A...1G}
{Gaia Collaboration} et~al., 2016, \mn@doi [\aap]
  {10.1051/0004-6361/201629272}, \href
  {https://ui.adsabs.harvard.edu/abs/2016A&A...595A...1G} {595, A1}

\bibitem[\protect\citeauthoryear{{Gaia Collaboration} et~al.,}{{Gaia
  Collaboration} et~al.}{2021}]{2021A&A...649A...6G}
{Gaia Collaboration} et~al., 2021, \mn@doi [\aap]
  {10.1051/0004-6361/202039498}, \href
  {https://ui.adsabs.harvard.edu/abs/2021A&A...649A...6G} {649, A6}

\bibitem[\protect\citeauthoryear{{Gaia Collaboration} et~al.,}{{Gaia
  Collaboration} et~al.}{2023a}]{2023A&A...674A...1G}
{Gaia Collaboration} et~al., 2023a, \mn@doi [\aap]
  {10.1051/0004-6361/202243940}, \href
  {https://ui.adsabs.harvard.edu/abs/2023A&A...674A...1G} {674, A1}

\bibitem[\protect\citeauthoryear{{Gaia Collaboration} et~al.,}{{Gaia
  Collaboration} et~al.}{2023b}]{2023A&A...674A..33G}
{Gaia Collaboration} et~al., 2023b, \mn@doi [\aap]
  {10.1051/0004-6361/202243709}, \href
  {https://ui.adsabs.harvard.edu/abs/2023A&A...674A..33G} {674, A33}

\bibitem[\protect\citeauthoryear{{Gaia Collaboration} et~al.,}{{Gaia
  Collaboration} et~al.}{2023c}]{2023A&A...674A..39G}
{Gaia Collaboration} et~al., 2023c, \mn@doi [\aap]
  {10.1051/0004-6361/202243800}, \href
  {https://ui.adsabs.harvard.edu/abs/2023A&A...674A..39G} {674, A39}

\bibitem[\protect\citeauthoryear{{G{\'a}lvez-Ortiz} et~al.,}{{G{\'a}lvez-Ortiz}
  et~al.}{2014}]{2014MNRAS.439.3890G}
{G{\'a}lvez-Ortiz} M.~C.,  et~al., 2014, \mn@doi [\mnras]
  {10.1093/mnras/stu241}, \href
  {https://ui.adsabs.harvard.edu/abs/2014MNRAS.439.3890G} {439, 3890}

\bibitem[\protect\citeauthoryear{{Geballe} et~al.,}{{Geballe}
  et~al.}{2002}]{2002ApJ...564..466G}
{Geballe} T.~R.,  et~al., 2002, \mn@doi [\apj] {10.1086/324078}, \href
  {https://ui.adsabs.harvard.edu/abs/2002ApJ...564..466G} {564, 466}

\bibitem[\protect\citeauthoryear{{Gizis}}{{Gizis}}{1997}]{1997AJ....113..806G}
{Gizis} J.~E.,  1997, \mn@doi [\aj] {10.1086/118302}, \href
  {https://ui.adsabs.harvard.edu/abs/1997AJ....113..806G} {113, 806}

\bibitem[\protect\citeauthoryear{{Gizis}}{{Gizis}}{2002}]{2002ApJ...575..484G}
{Gizis} J.~E.,  2002, \mn@doi [\apj] {10.1086/341259}, \href
  {https://ui.adsabs.harvard.edu/abs/2002ApJ...575..484G} {575, 484}

\bibitem[\protect\citeauthoryear{{Gizis} \& {Reid}}{{Gizis} \&
  {Reid}}{1999}]{1999AJ....117..508G}
{Gizis} J.~E.,  {Reid} I.~N.,  1999, \mn@doi [\aj] {10.1086/300709}, \href
  {https://ui.adsabs.harvard.edu/abs/1999AJ....117..508G} {117, 508}

\bibitem[\protect\citeauthoryear{{Gizis}, {Monet}, {Reid}, {Kirkpatrick},
  {Liebert}  \& {Williams}}{{Gizis} et~al.}{2000}]{2000AJ....120.1085G}
{Gizis} J.~E.,  {Monet} D.~G.,  {Reid} I.~N.,  {Kirkpatrick} J.~D.,  {Liebert}
  J.,   {Williams} R.~J.,  2000, \mn@doi [\aj] {10.1086/301456}, \href
  {https://ui.adsabs.harvard.edu/abs/2000AJ....120.1085G} {120, 1085}

\bibitem[\protect\citeauthoryear{{Gliese} \& {Jahrei{\ss}}}{{Gliese} \&
  {Jahrei{\ss}}}{1991}]{1991NSC3..C......0G}
{Gliese} W.,  {Jahrei{\ss}} H.,  1991, {Preliminary Version of the Third
  Catalogue of Nearby Stars}, On: The Astronomical Data Center CD-ROM: Selected
  Astronomical Catalogs, Vol. I; L.E. Brotzmann, S.E. Gesser (eds.),
  NASA/Astronomical Data Center, Goddard Space Flight Center, Greenbelt, MD

\bibitem[\protect\citeauthoryear{Harris et~al.,}{Harris
  et~al.}{2020}]{harris2020array}
Harris C.~R.,  et~al., 2020, \mn@doi [Nature] {10.1038/s41586-020-2649-2}, 585,
  357

\bibitem[\protect\citeauthoryear{{Hawley} et~al.,}{{Hawley}
  et~al.}{2002}]{2002AJ....123.3409H}
{Hawley} S.~L.,  et~al., 2002, \mn@doi [\aj] {10.1086/340697}, \href
  {https://ui.adsabs.harvard.edu/abs/2002AJ....123.3409H} {123, 3409}

\bibitem[\protect\citeauthoryear{{Hellemans}}{{Hellemans}}{1998}]{1998Sci...282.1240H}
{Hellemans} A.,  1998, \mn@doi [Science] {10.1126/science.282.5392.1240a},
  \href {https://ui.adsabs.harvard.edu/abs/1998Sci...282.1240H} {282, 1240}

\bibitem[\protect\citeauthoryear{Hunter}{Hunter}{2007}]{Hunter:2007}
Hunter J.~D.,  2007, \mn@doi [Computing in Science \& Engineering]
  {10.1109/MCSE.2007.55}, 9, 90

\bibitem[\protect\citeauthoryear{{Johnson} \& {Soderblom}}{{Johnson} \&
  {Soderblom}}{1987}]{1987AJ.....93..864J}
{Johnson} D. R.~H.,  {Soderblom} D.~R.,  1987, \mn@doi [\aj] {10.1086/114370},
  \href {https://ui.adsabs.harvard.edu/abs/1987AJ.....93..864J} {93, 864}

\bibitem[\protect\citeauthoryear{{Katz} et~al.,}{{Katz}
  et~al.}{2023}]{2023A&A...674A...5K}
{Katz} D.,  et~al., 2023, \mn@doi [\aap] {10.1051/0004-6361/202244220}, \href
  {https://ui.adsabs.harvard.edu/abs/2023A&A...674A...5K} {674, A5}

\bibitem[\protect\citeauthoryear{{Kellogg}, {Metchev}, {Miles-P{\'a}ez}  \&
  {Tannock}}{{Kellogg} et~al.}{2017}]{2017AJ....154..112K}
{Kellogg} K.,  {Metchev} S.,  {Miles-P{\'a}ez} P.~A.,   {Tannock} M.~E.,  2017,
  \mn@doi [\aj] {10.3847/1538-3881/aa83b0}, \href
  {https://ui.adsabs.harvard.edu/abs/2017AJ....154..112K} {154, 112}

\bibitem[\protect\citeauthoryear{{Kirkpatrick}}{{Kirkpatrick}}{2005}]{2005ARA&A..43..195K}
{Kirkpatrick} J.~D.,  2005, \mn@doi [\araa]
  {10.1146/annurev.astro.42.053102.134017}, \href
  {https://ui.adsabs.harvard.edu/abs/2005ARA&A..43..195K} {43, 195}

\bibitem[\protect\citeauthoryear{{Kirkpatrick} et~al.,}{{Kirkpatrick}
  et~al.}{1999}]{1999ApJ...519..802K}
{Kirkpatrick} J.~D.,  et~al., 1999, \mn@doi [\apj] {10.1086/307414}, \href
  {https://ui.adsabs.harvard.edu/abs/1999ApJ...519..802K} {519, 802}

\bibitem[\protect\citeauthoryear{{Kirkpatrick}, {Barman}, {Burgasser},
  {McGovern}, {McLean}, {Tinney}  \& {Lowrance}}{{Kirkpatrick}
  et~al.}{2006}]{2006ApJ...639.1120K}
{Kirkpatrick} J.~D.,  {Barman} T.~S.,  {Burgasser} A.~J.,  {McGovern} M.~R.,
  {McLean} I.~S.,  {Tinney} C.~G.,   {Lowrance} P.~J.,  2006, \mn@doi [\apj]
  {10.1086/499622}, \href
  {https://ui.adsabs.harvard.edu/abs/2006ApJ...639.1120K} {639, 1120}

\bibitem[\protect\citeauthoryear{{Kirkpatrick} et~al.,}{{Kirkpatrick}
  et~al.}{2008}]{2008ApJ...689.1295K}
{Kirkpatrick} J.~D.,  et~al., 2008, \mn@doi [\apj] {10.1086/592768}, \href
  {https://ui.adsabs.harvard.edu/abs/2008ApJ...689.1295K} {689, 1295}

\bibitem[\protect\citeauthoryear{{Kirkpatrick} et~al.,}{{Kirkpatrick}
  et~al.}{2010}]{2010ApJS..190..100K}
{Kirkpatrick} J.~D.,  et~al., 2010, \mn@doi [\apjs]
  {10.1088/0067-0049/190/1/100}, \href
  {https://ui.adsabs.harvard.edu/abs/2010ApJS..190..100K} {190, 100}

\bibitem[\protect\citeauthoryear{{Kirkpatrick} et~al.,}{{Kirkpatrick}
  et~al.}{2014}]{2014ApJ...783..122K}
{Kirkpatrick} J.~D.,  et~al., 2014, \mn@doi [\apj]
  {10.1088/0004-637X/783/2/122}, \href
  {https://ui.adsabs.harvard.edu/abs/2014ApJ...783..122K} {783, 122}

\bibitem[\protect\citeauthoryear{{Kirkpatrick} et~al.,}{{Kirkpatrick}
  et~al.}{2016}]{2016ApJS..224...36K}
{Kirkpatrick} J.~D.,  et~al., 2016, \mn@doi [\apjs]
  {10.3847/0067-0049/224/2/36}, \href
  {https://ui.adsabs.harvard.edu/abs/2016ApJS..224...36K} {224, 36}

\bibitem[\protect\citeauthoryear{{Kirkpatrick} et~al.,}{{Kirkpatrick}
  et~al.}{2021}]{2021ApJS..253....7K}
{Kirkpatrick} J.~D.,  et~al., 2021, \mn@doi [\apjs] {10.3847/1538-4365/abd107},
  \href {https://ui.adsabs.harvard.edu/abs/2021ApJS..253....7K} {253, 7}

\bibitem[\protect\citeauthoryear{{Koppelman}, {Helmi}  \&
  {Veljanoski}}{{Koppelman} et~al.}{2018}]{2018ApJ...860L..11K}
{Koppelman} H.,  {Helmi} A.,   {Veljanoski} J.,  2018, \mn@doi [\apjl]
  {10.3847/2041-8213/aac882}, \href
  {https://ui.adsabs.harvard.edu/abs/2018ApJ...860L..11K} {860, L11}

\bibitem[\protect\citeauthoryear{{Kounkel} et~al.,}{{Kounkel}
  et~al.}{2019}]{2019AJ....157..196K}
{Kounkel} M.,  et~al., 2019, \mn@doi [\aj] {10.3847/1538-3881/ab13b1}, \href
  {https://ui.adsabs.harvard.edu/abs/2019AJ....157..196K} {157, 196}

\bibitem[\protect\citeauthoryear{{Kraus} \& {Hillenbrand}}{{Kraus} \&
  {Hillenbrand}}{2007}]{2007ApJ...662..413K}
{Kraus} A.~L.,  {Hillenbrand} L.~A.,  2007, \mn@doi [\apj] {10.1086/516835},
  \href {https://ui.adsabs.harvard.edu/abs/2007ApJ...662..413K} {662, 413}

\bibitem[\protect\citeauthoryear{{Kraus}, {Herczeg}, {Rizzuto}, {Mann},
  {Slesnick}, {Carpenter}, {Hillenbrand}  \& {Mamajek}}{{Kraus}
  et~al.}{2017}]{2017ApJ...838..150K}
{Kraus} A.~L.,  {Herczeg} G.~J.,  {Rizzuto} A.~C.,  {Mann} A.~W.,  {Slesnick}
  C.~L.,  {Carpenter} J.~M.,  {Hillenbrand} L.~A.,   {Mamajek} E.~E.,  2017,
  \mn@doi [\apj] {10.3847/1538-4357/aa62a0}, \href
  {https://ui.adsabs.harvard.edu/abs/2017ApJ...838..150K} {838, 150}

\bibitem[\protect\citeauthoryear{{Leggett}}{{Leggett}}{1992}]{1992ApJS...82..351L}
{Leggett} S.~K.,  1992, \mn@doi [\apjs] {10.1086/191720}, \href
  {https://ui.adsabs.harvard.edu/abs/1992ApJS...82..351L} {82, 351}

\bibitem[\protect\citeauthoryear{{Leggett} \& {Hawkins}}{{Leggett} \&
  {Hawkins}}{1989}]{1989MNRAS.238..145L}
{Leggett} S.~K.,  {Hawkins} M.~R.~S.,  1989, \mn@doi [\mnras]
  {10.1093/mnras/238.1.145}, \href
  {https://ui.adsabs.harvard.edu/abs/1989MNRAS.238..145L} {238, 145}

\bibitem[\protect\citeauthoryear{{L{\'e}pine}}{{L{\'e}pine}}{2008}]{2008AJ....135.2177L}
{L{\'e}pine} S.,  2008, \mn@doi [\aj] {10.1088/0004-6256/135/6/2177}, \href
  {https://ui.adsabs.harvard.edu/abs/2008AJ....135.2177L} {135, 2177}

\bibitem[\protect\citeauthoryear{{L{\'e}pine}, {Shara}  \& {Rich}}{{L{\'e}pine}
  et~al.}{2002a}]{2002AJ....124.1190L}
{L{\'e}pine} S.,  {Shara} M.~M.,   {Rich} R.~M.,  2002a, \mn@doi [\aj]
  {10.1086/341783}, \href
  {https://ui.adsabs.harvard.edu/abs/2002AJ....124.1190L} {124, 1190}

\bibitem[\protect\citeauthoryear{{L{\'e}pine}, {Rich}, {Neill}, {Caulet}  \&
  {Shara}}{{L{\'e}pine} et~al.}{2002b}]{2002ApJ...581L..47L}
{L{\'e}pine} S.,  {Rich} R.~M.,  {Neill} J.~D.,  {Caulet} A.,   {Shara} M.~M.,
  2002b, \mn@doi [\apjl] {10.1086/345943}, \href
  {https://ui.adsabs.harvard.edu/abs/2002ApJ...581L..47L} {581, L47}

\bibitem[\protect\citeauthoryear{{L{\'e}pine}, {Rich}  \& {Shara}}{{L{\'e}pine}
  et~al.}{2003}]{2003ApJ...591L..49L}
{L{\'e}pine} S.,  {Rich} R.~M.,   {Shara} M.~M.,  2003, \mn@doi [\apjl]
  {10.1086/377069}, \href
  {https://ui.adsabs.harvard.edu/abs/2003ApJ...591L..49L} {591, L49}

\bibitem[\protect\citeauthoryear{{Liebert}, {Dahn}, {Gresham}  \&
  {Strittmatter}}{{Liebert} et~al.}{1979}]{1979ApJ...233..226L}
{Liebert} J.,  {Dahn} C.~C.,  {Gresham} M.,   {Strittmatter} P.~A.,  1979,
  \mn@doi [\apj] {10.1086/157384}, \href
  {https://ui.adsabs.harvard.edu/abs/1979ApJ...233..226L} {233, 226}

\bibitem[\protect\citeauthoryear{{Lodieu}, {Espinoza Contreras}, {Zapatero
  Osorio}, {Solano}, {Aberasturi}  \& {Mart{\'\i}n}}{{Lodieu}
  et~al.}{2012}]{2012A&A...542A.105L}
{Lodieu} N.,  {Espinoza Contreras} M.,  {Zapatero Osorio} M.~R.,  {Solano} E.,
  {Aberasturi} M.,   {Mart{\'\i}n} E.~L.,  2012, \mn@doi [\aap]
  {10.1051/0004-6361/201118717}, \href
  {https://ui.adsabs.harvard.edu/abs/2012A&A...542A.105L} {542, A105}

\bibitem[\protect\citeauthoryear{{Lodieu}, {Espinoza Contreras}, {Zapatero
  Osorio}, {Solano}, {Aberasturi}, {Mart{\'\i}n}  \& {Rodrigo}}{{Lodieu}
  et~al.}{2017}]{2017A&A...598A..92L}
{Lodieu} N.,  {Espinoza Contreras} M.,  {Zapatero Osorio} M.~R.,  {Solano} E.,
  {Aberasturi} M.,  {Mart{\'\i}n} E.~L.,   {Rodrigo} C.,  2017, \mn@doi [\aap]
  {10.1051/0004-6361/201629410}, \href
  {https://ui.adsabs.harvard.edu/abs/2017A&A...598A..92L} {598, A92}

\bibitem[\protect\citeauthoryear{{Looper}, {Burgasser}, {Kirkpatrick}  \&
  {Swift}}{{Looper} et~al.}{2007}]{2007ApJ...669L..97L}
{Looper} D.~L.,  {Burgasser} A.~J.,  {Kirkpatrick} J.~D.,   {Swift} B.~J.,
  2007, \mn@doi [\apjl] {10.1086/523812}, \href
  {https://ui.adsabs.harvard.edu/abs/2007ApJ...669L..97L} {669, L97}

\bibitem[\protect\citeauthoryear{{Luhman}}{{Luhman}}{2014a}]{2014ApJ...781....4L}
{Luhman} K.~L.,  2014a, \mn@doi [\apj] {10.1088/0004-637X/781/1/4}, \href
  {https://ui.adsabs.harvard.edu/abs/2014ApJ...781....4L} {781, 4}

\bibitem[\protect\citeauthoryear{{Luhman}}{{Luhman}}{2014b}]{2014ApJ...786L..18L}
{Luhman} K.~L.,  2014b, \mn@doi [\apjl] {10.1088/2041-8205/786/2/L18}, \href
  {https://ui.adsabs.harvard.edu/abs/2014ApJ...786L..18L} {786, L18}

\bibitem[\protect\citeauthoryear{{Luhman}}{{Luhman}}{2018}]{2018AJ....156..271L}
{Luhman} K.~L.,  2018, \mn@doi [\aj] {10.3847/1538-3881/aae831}, \href
  {https://ui.adsabs.harvard.edu/abs/2018AJ....156..271L} {156, 271}

\bibitem[\protect\citeauthoryear{{Luhman} \& {Sheppard}}{{Luhman} \&
  {Sheppard}}{2014}]{2014ApJ...787..126L}
{Luhman} K.~L.,  {Sheppard} S.~S.,  2014, \mn@doi [\apj]
  {10.1088/0004-637X/787/2/126}, \href
  {https://ui.adsabs.harvard.edu/abs/2014ApJ...787..126L} {787, 126}

\bibitem[\protect\citeauthoryear{{Luhman}, {Brice{\~n}o}, {Stauffer},
  {Hartmann}, {Barrado y Navascu{\'e}s}  \& {Caldwell}}{{Luhman}
  et~al.}{2003}]{2003ApJ...590..348L}
{Luhman} K.~L.,  {Brice{\~n}o} C.,  {Stauffer} J.~R.,  {Hartmann} L.,  {Barrado
  y Navascu{\'e}s} D.,   {Caldwell} N.,  2003, \mn@doi [\apj] {10.1086/374983},
  \href {https://ui.adsabs.harvard.edu/abs/2003ApJ...590..348L} {590, 348}

\bibitem[\protect\citeauthoryear{{Luhman}, {Whitney}, {Meade}, {Babler},
  {Indebetouw}, {Bracker}  \& {Churchwell}}{{Luhman}
  et~al.}{2006}]{2006ApJ...647.1180L}
{Luhman} K.~L.,  {Whitney} B.~A.,  {Meade} M.~R.,  {Babler} B.~L.,
  {Indebetouw} R.,  {Bracker} S.,   {Churchwell} E.~B.,  2006, \mn@doi [\apj]
  {10.1086/505572}, \href
  {https://ui.adsabs.harvard.edu/abs/2006ApJ...647.1180L} {647, 1180}

\bibitem[\protect\citeauthoryear{{Luhman}, {Mamajek}, {Allen}  \&
  {Cruz}}{{Luhman} et~al.}{2009}]{2009ApJ...703..399L}
{Luhman} K.~L.,  {Mamajek} E.~E.,  {Allen} P.~R.,   {Cruz} K.~L.,  2009,
  \mn@doi [\apj] {10.1088/0004-637X/703/1/399}, \href
  {https://ui.adsabs.harvard.edu/abs/2009ApJ...703..399L} {703, 399}

\bibitem[\protect\citeauthoryear{{Luhman}, {Allen}, {Espaillat}, {Hartmann}  \&
  {Calvet}}{{Luhman} et~al.}{2010}]{2010ApJS..186..111L}
{Luhman} K.~L.,  {Allen} P.~R.,  {Espaillat} C.,  {Hartmann} L.,   {Calvet} N.,
   2010, \mn@doi [\apjs] {10.1088/0067-0049/186/1/111}, \href
  {https://ui.adsabs.harvard.edu/abs/2010ApJS..186..111L} {186, 111}

\bibitem[\protect\citeauthoryear{{Luhman}, {Esplin}  \& {Loutrel}}{{Luhman}
  et~al.}{2016}]{2016ApJ...827...52L}
{Luhman} K.~L.,  {Esplin} T.~L.,   {Loutrel} N.~P.,  2016, \mn@doi [\apj]
  {10.3847/0004-637X/827/1/52}, \href
  {https://ui.adsabs.harvard.edu/abs/2016ApJ...827...52L} {827, 52}

\bibitem[\protect\citeauthoryear{{Luhman}, {Mamajek}, {Shukla}  \&
  {Loutrel}}{{Luhman} et~al.}{2017}]{2017AJ....153...46L}
{Luhman} K.~L.,  {Mamajek} E.~E.,  {Shukla} S.~J.,   {Loutrel} N.~P.,  2017,
  \mn@doi [\aj] {10.3847/1538-3881/153/1/46}, \href
  {https://ui.adsabs.harvard.edu/abs/2017AJ....153...46L} {153, 46}

\bibitem[\protect\citeauthoryear{{Luhman}, {Herrmann}, {Mamajek}, {Esplin}  \&
  {Pecaut}}{{Luhman} et~al.}{2018}]{2018AJ....156...76L}
{Luhman} K.~L.,  {Herrmann} K.~A.,  {Mamajek} E.~E.,  {Esplin} T.~L.,
  {Pecaut} M.~J.,  2018, \mn@doi [\aj] {10.3847/1538-3881/aacc6d}, \href
  {https://ui.adsabs.harvard.edu/abs/2018AJ....156...76L} {156, 76}

\bibitem[\protect\citeauthoryear{{Magazz{\`u}}, {Dougados}, {Licandro},
  {Mart{\'\i}n}, {Magnier}  \& {M{\'e}nard}}{{Magazz{\`u}}
  et~al.}{2003}]{2003IAUS..211...75M}
{Magazz{\`u}} A.,  {Dougados} C.,  {Licandro} J.,  {Mart{\'\i}n} E.~L.,
  {Magnier} E.~A.,   {M{\'e}nard} F.,  2003, in {Mart{\'\i}n} E.,  ed.,
  Proceedings of IAU Symposium Vol. 211, Brown Dwarfs. p.~75

\bibitem[\protect\citeauthoryear{{Marocco} et~al.,}{{Marocco}
  et~al.}{2013}]{2013AJ....146..161M}
{Marocco} F.,  et~al., 2013, \mn@doi [\aj] {10.1088/0004-6256/146/6/161}, \href
  {https://ui.adsabs.harvard.edu/abs/2013AJ....146..161M} {146, 161}

\bibitem[\protect\citeauthoryear{{Marocco} et~al.,}{{Marocco}
  et~al.}{2017}]{2017MNRAS.470.4885M}
{Marocco} F.,  et~al., 2017, \mn@doi [\mnras] {10.1093/mnras/stx1500}, \href
  {https://ui.adsabs.harvard.edu/abs/2017MNRAS.470.4885M} {470, 4885}

\bibitem[\protect\citeauthoryear{{Marocco} et~al.,}{{Marocco}
  et~al.}{2020}]{2020MNRAS.494.4891M}
{Marocco} F.,  et~al., 2020, \mn@doi [\mnras] {10.1093/mnras/staa1007}, \href
  {https://ui.adsabs.harvard.edu/abs/2020MNRAS.494.4891M} {494, 4891}

\bibitem[\protect\citeauthoryear{{Mart{\'\i}n}, {Delfosse}, {Basri}, {Goldman},
  {Forveille}  \& {Zapatero Osorio}}{{Mart{\'\i}n}
  et~al.}{1999}]{1999AJ....118.2466M}
{Mart{\'\i}n} E.~L.,  {Delfosse} X.,  {Basri} G.,  {Goldman} B.,  {Forveille}
  T.,   {Zapatero Osorio} M.~R.,  1999, \mn@doi [\aj] {10.1086/301107}, \href
  {https://ui.adsabs.harvard.edu/abs/1999AJ....118.2466M} {118, 2466}

\bibitem[\protect\citeauthoryear{{M{\'e}nard}, {Delfosse}  \&
  {Monin}}{{M{\'e}nard} et~al.}{2002}]{2002A&A...396L..35M}
{M{\'e}nard} F.,  {Delfosse} X.,   {Monin} J.~L.,  2002, \mn@doi [\aap]
  {10.1051/0004-6361:20021657}, \href
  {https://ui.adsabs.harvard.edu/abs/2002A&A...396L..35M} {396, L35}

\bibitem[\protect\citeauthoryear{Meurer et~al.,}{Meurer
  et~al.}{2017}]{10.7717/peerj-cs.103}
Meurer A.,  et~al., 2017, \mn@doi [PeerJ Computer Science]
  {10.7717/peerj-cs.103}, 3, e103

\bibitem[\protect\citeauthoryear{{Montegriffo} et~al.,}{{Montegriffo}
  et~al.}{2023}]{2023A&A...674A...3M}
{Montegriffo} P.,  et~al., 2023, \mn@doi [\aap] {10.1051/0004-6361/202243880},
  \href {https://ui.adsabs.harvard.edu/abs/2023A&A...674A...3M} {674, A3}

\bibitem[\protect\citeauthoryear{{Nissen} \& {Schuster}}{{Nissen} \&
  {Schuster}}{2010}]{2010A&A...511L..10N}
{Nissen} P.~E.,  {Schuster} W.~J.,  2010, \mn@doi [\aap]
  {10.1051/0004-6361/200913877}, \href
  {https://ui.adsabs.harvard.edu/abs/2010A&A...511L..10N} {511, L10}

\bibitem[\protect\citeauthoryear{{Pancino} et~al.,}{{Pancino}
  et~al.}{2012}]{2012MNRAS.426.1767P}
{Pancino} E.,  et~al., 2012, \mn@doi [\mnras]
  {10.1111/j.1365-2966.2012.21766.x}, \href
  {https://ui.adsabs.harvard.edu/abs/2012MNRAS.426.1767P} {426, 1767}

\bibitem[\protect\citeauthoryear{{Phan-Bao} et~al.,}{{Phan-Bao}
  et~al.}{2003}]{2003A&A...401..959P}
{Phan-Bao} N.,  et~al., 2003, \mn@doi [\aap] {10.1051/0004-6361:20030188},
  \href {https://ui.adsabs.harvard.edu/abs/2003A&A...401..959P} {401, 959}

\bibitem[\protect\citeauthoryear{{Rebull} et~al.,}{{Rebull}
  et~al.}{2010}]{2010ApJS..186..259R}
{Rebull} L.~M.,  et~al., 2010, \mn@doi [\apjs] {10.1088/0067-0049/186/2/259},
  \href {https://ui.adsabs.harvard.edu/abs/2010ApJS..186..259R} {186, 259}

\bibitem[\protect\citeauthoryear{{Rebull}, {Stauffer}, {Cody}, {Hillenbrand},
  {Bouvier}, {Roggero}  \& {David}}{{Rebull}
  et~al.}{2020}]{2020AJ....159..273R}
{Rebull} L.~M.,  {Stauffer} J.~R.,  {Cody} A.~M.,  {Hillenbrand} L.~A.,
  {Bouvier} J.,  {Roggero} N.,   {David} T.~J.,  2020, \mn@doi [\aj]
  {10.3847/1538-3881/ab893c}, \href
  {https://ui.adsabs.harvard.edu/abs/2020AJ....159..273R} {159, 273}

\bibitem[\protect\citeauthoryear{{Reid}, {Burgasser}, {Cruz}, {Kirkpatrick}  \&
  {Gizis}}{{Reid} et~al.}{2001}]{2001AJ....121.1710R}
{Reid} I.~N.,  {Burgasser} A.~J.,  {Cruz} K.~L.,  {Kirkpatrick} J.~D.,
  {Gizis} J.~E.,  2001, \mn@doi [\aj] {10.1086/319418}, \href
  {https://ui.adsabs.harvard.edu/abs/2001AJ....121.1710R} {121, 1710}

\bibitem[\protect\citeauthoryear{{Reid}, {Lewitus}, {Allen}, {Cruz}  \&
  {Burgasser}}{{Reid} et~al.}{2006}]{2006AJ....132..891R}
{Reid} I.~N.,  {Lewitus} E.,  {Allen} P.~R.,  {Cruz} K.~L.,   {Burgasser}
  A.~J.,  2006, \mn@doi [\aj] {10.1086/505626}, \href
  {https://ui.adsabs.harvard.edu/abs/2006AJ....132..891R} {132, 891}

\bibitem[\protect\citeauthoryear{{Reid}, {Cruz}, {Kirkpatrick}, {Allen},
  {Mungall}, {Liebert}, {Lowrance}  \& {Sweet}}{{Reid}
  et~al.}{2008}]{2008AJ....136.1290R}
{Reid} I.~N.,  {Cruz} K.~L.,  {Kirkpatrick} J.~D.,  {Allen} P.~R.,  {Mungall}
  F.,  {Liebert} J.,  {Lowrance} P.,   {Sweet} A.,  2008, \mn@doi [\aj]
  {10.1088/0004-6256/136/3/1290}, \href
  {https://ui.adsabs.harvard.edu/abs/2008AJ....136.1290R} {136, 1290}

\bibitem[\protect\citeauthoryear{{Reiners} \& {Basri}}{{Reiners} \&
  {Basri}}{2006}]{2006AJ....131.1806R}
{Reiners} A.,  {Basri} G.,  2006, \mn@doi [\aj] {10.1086/500298}, \href
  {https://ui.adsabs.harvard.edu/abs/2006AJ....131.1806R} {131, 1806}

\bibitem[\protect\citeauthoryear{{Reyl{\'e}}, {Jardine}, {Fouqu{\'e}},
  {Caballero}, {Smart}  \& {Sozzetti}}{{Reyl{\'e}}
  et~al.}{2021}]{2021A&A...650A.201R}
{Reyl{\'e}} C.,  {Jardine} K.,  {Fouqu{\'e}} P.,  {Caballero} J.~A.,  {Smart}
  R.~L.,   {Sozzetti} A.,  2021, \mn@doi [\aap] {10.1051/0004-6361/202140985},
  \href {https://ui.adsabs.harvard.edu/abs/2021A&A...650A.201R} {650, A201}

\bibitem[\protect\citeauthoryear{{Riello} et~al.,}{{Riello}
  et~al.}{2021}]{2021A&A...649A...3R}
{Riello} M.,  et~al., 2021, \mn@doi [\aap] {10.1051/0004-6361/202039587}, \href
  {https://ui.adsabs.harvard.edu/abs/2021A&A...649A...3R} {649, A3}

\bibitem[\protect\citeauthoryear{Ruz-Mieres}{Ruz-Mieres}{2022}]{daniela_ruz_mieres_2022_6674521}
Ruz-Mieres D.,  2022, gaia-dpci/GaiaXPy: GaiaXPy 1.1.4,
  \mn@doi{10.5281/zenodo.6674521}, \url
  {https://doi.org/10.5281/zenodo.6674521}

\bibitem[\protect\citeauthoryear{{Salim}, {L{\'e}pine}, {Rich}  \&
  {Shara}}{{Salim} et~al.}{2003}]{2003ApJ...586L.149S}
{Salim} S.,  {L{\'e}pine} S.,  {Rich} R.~M.,   {Shara} M.~M.,  2003, \mn@doi
  [\apjl] {10.1086/374794}, \href
  {https://ui.adsabs.harvard.edu/abs/2003ApJ...586L.149S} {586, L149}

\bibitem[\protect\citeauthoryear{{Sandage} \& {Fouts}}{{Sandage} \&
  {Fouts}}{1987}]{1987AJ.....93...74S}
{Sandage} A.,  {Fouts} G.,  1987, \mn@doi [\aj] {10.1086/114291}, \href
  {https://ui.adsabs.harvard.edu/abs/1987AJ.....93...74S} {93, 74}

\bibitem[\protect\citeauthoryear{{Sarro} et~al.,}{{Sarro}
  et~al.}{2023}]{2023A&A...669A.139S}
{Sarro} L.~M.,  et~al., 2023, \mn@doi [\aap] {10.1051/0004-6361/202244507},
  \href {https://ui.adsabs.harvard.edu/abs/2023A&A...669A.139S} {669, A139}

\bibitem[\protect\citeauthoryear{{Sartoretti} et~al.,}{{Sartoretti}
  et~al.}{2023}]{2023A&A...674A...6S}
{Sartoretti} P.,  et~al., 2023, \mn@doi [\aap] {10.1051/0004-6361/202243615},
  \href {https://ui.adsabs.harvard.edu/abs/2023A&A...674A...6S} {674, A6}

\bibitem[\protect\citeauthoryear{{Schmidt}, {Cruz}, {Bongiorno}, {Liebert}  \&
  {Reid}}{{Schmidt} et~al.}{2007}]{2007AJ....133.2258S}
{Schmidt} S.~J.,  {Cruz} K.~L.,  {Bongiorno} B.~J.,  {Liebert} J.,   {Reid}
  I.~N.,  2007, \mn@doi [\aj] {10.1086/512158}, \href
  {https://ui.adsabs.harvard.edu/abs/2007AJ....133.2258S} {133, 2258}

\bibitem[\protect\citeauthoryear{{Schneider}, {Melis}, {Song}  \&
  {Zuckerman}}{{Schneider} et~al.}{2011}]{2011ApJ...743..109S}
{Schneider} A.,  {Melis} C.,  {Song} I.,   {Zuckerman} B.,  2011, \mn@doi
  [\apj] {10.1088/0004-637X/743/2/109}, \href
  {https://ui.adsabs.harvard.edu/abs/2011ApJ...743..109S} {743, 109}

\bibitem[\protect\citeauthoryear{{Schneider}, {Greco}, {Cushing},
  {Kirkpatrick}, {Mainzer}, {Gelino}, {Fajardo-Acosta}  \& {Bauer}}{{Schneider}
  et~al.}{2016}]{2016ApJ...817..112S}
{Schneider} A.~C.,  {Greco} J.,  {Cushing} M.~C.,  {Kirkpatrick} J.~D.,
  {Mainzer} A.,  {Gelino} C.~R.,  {Fajardo-Acosta} S.~B.,   {Bauer} J.,  2016,
  \mn@doi [\apj] {10.3847/0004-637X/817/2/112}, \href
  {https://ui.adsabs.harvard.edu/abs/2016ApJ...817..112S} {817, 112}

\bibitem[\protect\citeauthoryear{{Scholz}, {Lodieu}  \& {McCaughrean}}{{Scholz}
  et~al.}{2004}]{2004A&A...428L..25S}
{Scholz} R.~D.,  {Lodieu} N.,   {McCaughrean} M.~J.,  2004, \mn@doi [\aap]
  {10.1051/0004-6361:200400098}, \href
  {https://ui.adsabs.harvard.edu/abs/2004A&A...428L..25S} {428, L25}

\bibitem[\protect\citeauthoryear{{Scholz}, {McCaughrean}, {Zinnecker}  \&
  {Lodieu}}{{Scholz} et~al.}{2005}]{2005A&A...430L..49S}
{Scholz} R.~D.,  {McCaughrean} M.~J.,  {Zinnecker} H.,   {Lodieu} N.,  2005,
  \mn@doi [\aap] {10.1051/0004-6361:200400121}, \href
  {https://ui.adsabs.harvard.edu/abs/2005A&A...430L..49S} {430, L49}

\bibitem[\protect\citeauthoryear{{Sch{\"o}nrich} \& {Binney}}{{Sch{\"o}nrich}
  \& {Binney}}{2009}]{2009MNRAS.399.1145S}
{Sch{\"o}nrich} R.,  {Binney} J.,  2009, \mn@doi [\mnras]
  {10.1111/j.1365-2966.2009.15365.x}, \href
  {https://ui.adsabs.harvard.edu/abs/2009MNRAS.399.1145S} {399, 1145}

\bibitem[\protect\citeauthoryear{{Skrutskie} et~al.,}{{Skrutskie}
  et~al.}{2006}]{2006AJ....131.1163S}
{Skrutskie} M.~F.,  et~al., 2006, \mn@doi [\aj] {10.1086/498708}, \href
  {https://ui.adsabs.harvard.edu/abs/2006AJ....131.1163S} {131, 1163}

\bibitem[\protect\citeauthoryear{{Smart}, {Marocco}, {Caballero}, {Jones},
  {Barrado}, {Beam{\'{\i}}n}, {Pinfield}  \& {Sarro}}{{Smart}
  et~al.}{2017}]{2017MNRAS.469..401S}
{Smart} R.~L.,  {Marocco} F.,  {Caballero} J.~A.,  {Jones} H.~R.~A.,  {Barrado}
  D.,  {Beam{\'{\i}}n} J.~C.,  {Pinfield} D.~J.,   {Sarro} L.~M.,  2017,
  \mn@doi [\mnras] {10.1093/mnras/stx800}, \href
  {http://cdsads.u-strasbg.fr/abs/2017MNRAS.469..401S} {469, 401}

\bibitem[\protect\citeauthoryear{{Smart}, {Marocco}, {Sarro}, {Barrado},
  {Beam{\'\i}n}, {Caballero}  \& {Jones}}{{Smart}
  et~al.}{2019}]{2019MNRAS.485.4423S}
{Smart} R.~L.,  {Marocco} F.,  {Sarro} L.~M.,  {Barrado} D.,  {Beam{\'\i}n}
  J.~C.,  {Caballero} J.~A.,   {Jones} H.~R.~A.,  2019, \mn@doi [\mnras]
  {10.1093/mnras/stz678}, \href
  {https://ui.adsabs.harvard.edu/abs/2019MNRAS.485.4423S} {485, 4423}

\bibitem[\protect\citeauthoryear{{Stephens} et~al.,}{{Stephens}
  et~al.}{2009}]{2009ApJ...702..154S}
{Stephens} D.~C.,  et~al., 2009, \mn@doi [\apj] {10.1088/0004-637X/702/1/154},
  \href {https://ui.adsabs.harvard.edu/abs/2009ApJ...702..154S} {702, 154}

\bibitem[\protect\citeauthoryear{{Tinney}}{{Tinney}}{1993}]{1993ApJ...414..279T}
{Tinney} C.~G.,  1993, \mn@doi [\apj] {10.1086/173075}, \href
  {https://ui.adsabs.harvard.edu/abs/1993ApJ...414..279T} {414, 279}

\bibitem[\protect\citeauthoryear{{Tinney} \& {Reid}}{{Tinney} \&
  {Reid}}{1998}]{1998MNRAS.301.1031T}
{Tinney} C.~G.,  {Reid} I.~N.,  1998, \mn@doi [\mnras]
  {10.1046/j.1365-8711.1998.02079.x}, \href
  {https://ui.adsabs.harvard.edu/abs/1998MNRAS.301.1031T} {301, 1031}

\bibitem[\protect\citeauthoryear{{Venn}, {Irwin}, {Shetrone}, {Tout}, {Hill}
  \& {Tolstoy}}{{Venn} et~al.}{2004}]{2004AJ....128.1177V}
{Venn} K.~A.,  {Irwin} M.,  {Shetrone} M.~D.,  {Tout} C.~A.,  {Hill} V.,
  {Tolstoy} E.,  2004, \mn@doi [\aj] {10.1086/422734}, \href
  {https://ui.adsabs.harvard.edu/abs/2004AJ....128.1177V} {128, 1177}

\bibitem[\protect\citeauthoryear{Virtanen et~al.,}{Virtanen
  et~al.}{2020}]{2020SciPy-NMeth}
Virtanen P.,  et~al., 2020, \mn@doi [Nature Methods]
  {10.1038/s41592-019-0686-2}, \href {https://rdcu.be/b08Wh} {17, 261}

\bibitem[\protect\citeauthoryear{{Wenger} et~al.,}{{Wenger}
  et~al.}{2000}]{2000A&AS..143....9W}
{Wenger} M.,  et~al., 2000, \mn@doi [\aaps] {10.1051/aas:2000332}, \href
  {https://ui.adsabs.harvard.edu/abs/2000A&AS..143....9W} {143, 9}

\bibitem[\protect\citeauthoryear{{W}es {M}c{K}inney}{{W}es
  {M}c{K}inney}{2010}]{mckinney-proc-scipy-2010}
{W}es {M}c{K}inney 2010, in {S}t\'efan van~der {W}alt {J}arrod {M}illman eds,
  {P}roceedings of the 9th {P}ython in {S}cience {C}onference. pp 56 -- 61,
  \mn@doi{10.25080/Majora-92bf1922-00a}

\bibitem[\protect\citeauthoryear{{West}, {Hawley}, {Bochanski}, {Covey},
  {Reid}, {Dhital}, {Hilton}  \& {Masuda}}{{West}
  et~al.}{2008}]{2008AJ....135..785W}
{West} A.~A.,  {Hawley} S.~L.,  {Bochanski} J.~J.,  {Covey} K.~R.,  {Reid}
  I.~N.,  {Dhital} S.,  {Hilton} E.~J.,   {Masuda} M.,  2008, \mn@doi [\aj]
  {10.1088/0004-6256/135/3/785}, \href
  {https://ui.adsabs.harvard.edu/abs/2008AJ....135..785W} {135, 785}

\bibitem[\protect\citeauthoryear{{Wilson}, {Miller}, {Gizis}, {Skrutskie},
  {Houck}, {Kirkpatrick}, {Burgasser}  \& {Monet}}{{Wilson}
  et~al.}{2003}]{2003IAUS..211..197W}
{Wilson} J.~C.,  {Miller} N.~A.,  {Gizis} J.~E.,  {Skrutskie} M.~F.,  {Houck}
  J.~R.,  {Kirkpatrick} J.~D.,  {Burgasser} A.~J.,   {Monet} D.~G.,  2003, in
  {Mart{\'\i}n} E.,  ed.,  Proceedings of IAU Symposium Vol. 211, Brown Dwarfs.
  p.~197

\bibitem[\protect\citeauthoryear{{Winters} et~al.,}{{Winters}
  et~al.}{2015}]{2015AJ....149....5W}
{Winters} J.~G.,  et~al., 2015, \mn@doi [\aj] {10.1088/0004-6256/149/1/5},
  \href {https://ui.adsabs.harvard.edu/abs/2015AJ....149....5W} {149, 5}

\bibitem[\protect\citeauthoryear{{Yao}, {Ji}, {Koposov}  \& {Limberg}}{{Yao}
  et~al.}{2023}]{2023arXiv230317676Y}
{Yao} Y.,  {Ji} A.~P.,  {Koposov} S.~E.,   {Limberg} G.,  2023, \mn@doi [arXiv
  e-prints] {10.48550/arXiv.2303.17676}, \href
  {https://ui.adsabs.harvard.edu/abs/2023arXiv230317676Y} {p. arXiv:2303.17676}

\bibitem[\protect\citeauthoryear{{Zhang} \& {Zhao}}{{Zhang} \&
  {Zhao}}{2006}]{2006A&A...449..127Z}
{Zhang} H.~W.,  {Zhao} G.,  2006, \mn@doi [\aap] {10.1051/0004-6361:20053770},
  \href {https://ui.adsabs.harvard.edu/abs/2006A&A...449..127Z} {449, 127}

\bibitem[\protect\citeauthoryear{{Zhang} et~al.,}{{Zhang}
  et~al.}{2017a}]{2017MNRAS.464.3040Z}
{Zhang} Z.~H.,  et~al., 2017a, \mn@doi [\mnras] {10.1093/mnras/stw2438}, \href
  {https://ui.adsabs.harvard.edu/abs/2017MNRAS.464.3040Z} {464, 3040}

\bibitem[\protect\citeauthoryear{{Zhang}, {Homeier}, {Pinfield}, {Lodieu},
  {Jones}, {Allard}  \& {Pavlenko}}{{Zhang}
  et~al.}{2017b}]{2017MNRAS.468..261Z}
{Zhang} Z.~H.,  {Homeier} D.,  {Pinfield} D.~J.,  {Lodieu} N.,  {Jones}
  H.~R.~A.,  {Allard} F.,   {Pavlenko} Y.~V.,  2017b, \mn@doi [\mnras]
  {10.1093/mnras/stx350}, \href
  {https://ui.adsabs.harvard.edu/abs/2017MNRAS.468..261Z} {468, 261}

\bibitem[\protect\citeauthoryear{{Zhang} et~al.,}{{Zhang}
  et~al.}{2018}]{2018ApJ...858...41Z}
{Zhang} Z.,  et~al., 2018, \mn@doi [\apj] {10.3847/1538-4357/aab269}, \href
  {https://ui.adsabs.harvard.edu/abs/2018ApJ...858...41Z} {858, 41}

\bibitem[\protect\citeauthoryear{{Zhang}, {Burgasser}  \& {Smith}}{{Zhang}
  et~al.}{2019}]{2019MNRAS.486.1840Z}
{Zhang} Z.~H.,  {Burgasser} A.~J.,   {Smith} L.~C.,  2019, \mn@doi [\mnras]
  {10.1093/mnras/stz659}, \href
  {https://ui.adsabs.harvard.edu/abs/2019MNRAS.486.1840Z} {486, 1840}

\bibitem[\protect\citeauthoryear{{Zhang}, {Green}  \& {Rix}}{{Zhang}
  et~al.}{2023}]{2023MNRAS.tmp.1869Z}
{Zhang} X.,  {Green} G.~M.,   {Rix} H.-W.,  2023, \mn@doi [\mnras]
  {10.1093/mnras/stad1941}, \href
  {https://ui.adsabs.harvard.edu/abs/2023MNRAS.tmp.1869Z} {}

\bibitem[\protect\citeauthoryear{da Costa-Luis et~al.,}{da~Costa-Luis
  et~al.}{2021}]{casper_da_costa_luis_2021_5517697}
da Costa-Luis C.,  et~al., 2021, {tqdm: A fast, Extensible Progress Bar for
  Python and CLI}, \mn@doi{10.5281/zenodo.5517697}, \url
  {https://doi.org/10.5281/zenodo.5517697}

\bibitem[\protect\citeauthoryear{pandas~development team}{pandas~development
  team}{2020}]{reback2020pandas}
pandas~development team T.,  2020, pandas-dev/pandas: Pandas,
  \mn@doi{10.5281/zenodo.3509134}, \url
  {https://doi.org/10.5281/zenodo.3509134}

\bibitem[\protect\citeauthoryear{van Rossum \& de Boer}{van Rossum \&
  de~Boer}{1991}]{python}
van Rossum G.,  de Boer J.,  1991, CWI Quarterly, 4, 283

\makeatother
\end{thebibliography}

\appendix
\section*{Appendix}

\onecolumn
\begin{longtable}{c cc c ll c}
\caption{
\label{tab:litobjects}
List of subdwarfs and young objects used to train our colour ratio.
Astrometry is from \gdrthree\ and the \teff\ values are those produced by the \espucd\ Apsis module and 
published as part of the Data Release.
}\\
\hline \hline
    \gdrthree  & $\alpha$ & $\delta$ & $\varpi$ & Object   & Spectral   & Teff \\
     Source ID & [hms]     & [dms]    & [mas]     & Name     & Type       & [K] \\
\hline
\endfirsthead
\hline
    \gdrthree  & $\alpha$ & $\delta$ & $\varpi$ & Object   & Spectral   & Teff \\
     Source ID & [hms]     & [dms]    & [mas]     & Name     & Type       & [K] \\
\hline
\endhead
\hline
\endfoot
\hline
\caption*{\hypertarget{litobjrefs}{References}: 1.~\cite{2014ApJ...784..126E}, 
2.~\cite{2017AJ....153...46L}, 
3.~\cite{2003ApJ...591L..49L}, 
4.~\cite{2006AJ....131.1806R}, 
5.~\cite{2015ApJ...798...73G}, 
6.~\cite{2015ApJS..219...33G}, 
7.~\cite{1991NSC3..C......0G}, 
8.~\cite{2012ApJ...752...56F}, 
9.~\cite{2018AJ....156...76L}, 
10.~\cite{2010ApJS..186..259R}, 
11.~\cite{2011ApJ...743..109S}, 
12.~\cite{2016AJ....151...46A}, 
13.~\cite{2000AJ....120..479A}, 
14.~\cite{1993ApJ...414..279T}, 
15.~\cite{2010ApJS..190..100K}, 
16.~\cite{2007A&A...468..163D}, 
17.~\cite{2004AJ....127.2856B}, 
18.~\cite{2014ApJ...794..143B}, 
19.~\cite{2011AJ....141...54A}, 
20.~\cite{2013AJ....146..161M}, 
21.~\cite{2002ApJ...581L..47L}, 
22.~\cite{2003IAUS..211...75M}, 
23.~\cite{2008AJ....136.1290R}, 
24.~\cite{2003A&A...401..959P}, 
25.~\cite{2008ApJ...689.1295K}, 
26.~\cite{2017AJ....154..112K}, 
27.~\cite{2000AJ....120.1085G}, 
28.~\cite{2012ApJS..201...19D}, 
29.~\cite{1979ApJ...233..226L}, 
30.~\cite{2005A&A...430L..49S}, 
31.~\cite{2017ApJS..228...18G}, 
32.~\cite{2003AJ....126.2421C}, 
33.~\cite{1989MNRAS.238..145L}, 
34.~\cite{2009ApJ...703..399L}, 
35.~\cite{2002ApJ...575..484G}, 
36.~\cite{2013ApJ...772...79A}, 
37.~\cite{2004A&A...428L..25S}, 
38.~\cite{2015AJ....149....5W}, 
39.~\cite{2006ApJ...649..862C}, 
40.~\cite{2016ApJ...827...52L}, 
41.~\cite{2014ApJ...785L..14G}, 
42.~\cite{2002A&A...396L..35M}, 
43.~\cite{2014MNRAS.439.3890G}, 
44.~\cite{2004ApJ...614L..73B}, 
45.~\cite{2002AJ....123.3409H}, 
46.~\cite{2007AJ....133..439C}, 
47.~\cite{2014ApJ...781....4L}, 
48.~\cite{2016ApJS..224...36K}, 
49.~\cite{2003IAUS..211..197W}, 
50.~\cite{2009AJ....137.3345C}, 
51.~\cite{2007ApJ...669L..97L}, 
52.~\cite{2008AJ....135..785W}, 
53.~\cite{2017MNRAS.464.3040Z}, 
54.~\cite{2014ApJ...783..122K}, 
55.~\cite{2006ApJ...639.1120K}, 
56.~\cite{1999A&A...351L...5E}, 
57.~\cite{2003AJ....126.1526B}, 
58.~\cite{2016ApJ...817..112S}, 
59.~\cite{2007AJ....133.2258S}, 
60.~\cite{2003ApJ...586L.149S}, 
61.~\cite{1998Sci...282.1240H}, 
62.~\cite{2006AJ....132..891R}}
\endlastfoot
    164802984685384320 & 4 15 41 & +29 15 07.6 & $6.5\pm0.1$ & 2MASS J04154131$+$2915078$\hyperlink{litobjrefs}{^{1}}$ & M8$\gamma$$\hyperlink{litobjrefs}{^{2}}$ & $2664\pm13$\\
    4406489184157821952 & 16 10 28 & -0 41 13.7 & $33.5\pm0.3$ & LSR J1610$-$0040$\hyperlink{litobjrefs}{^{3}}$ & d/sdM6$\hyperlink{litobjrefs}{^{4}}$ & $2651\pm11$\\
    152466120624336896 & 4 26 45 & +27 56 42.9 & $7.4\pm0.1$ & 2MASS J04264449$+$2756433$\hyperlink{litobjrefs}{^{1}}$ & M7$\gamma$$\hyperlink{litobjrefs}{^{2}}$ & $2674\pm19$\\
    3406128761895775872 & 4 44 02 & +16 21 32.1 & $6.9\pm0.1$ & 2MASS J04440164$+$1621324$\hyperlink{litobjrefs}{^{1}}$ & M7$\gamma$$\hyperlink{litobjrefs}{^{1}}$ & $2670\pm14$\\
    52039511681854208 & 4 10 28 & +20 51 50.5 & $7.7\pm0.4$ & 2MASS J04102834$+$2051507$\hyperlink{litobjrefs}{^{1}}$ & M7$\gamma$$\hyperlink{litobjrefs}{^{1}}$ & $2688\pm20$\\
    6412696995416769536 & 22 02 58 & -56 05 10.0 & $14.4\pm0.3$ & 2MASS J22025794$-$5605087$\hyperlink{litobjrefs}{^{5}}$ & M6.2$\gamma$$\hyperlink{litobjrefs}{^{6}}$ & $2322\pm27$\\
    3311992669430199168 & 4 22 14 & +15 30 52.6 & $3.5\pm0.1$ & Cl* Melotte   25     LH     190$\hyperlink{litobjrefs}{^{7}}$ & M6:$\gamma$$\hyperlink{litobjrefs}{^{8}}$ & $2527\pm19$\\
    6154629964132559104 & 12 57 45 & -36 35 43.4 & $12.3\pm0.2$ & 2MASS J12574463$-$3635431$\hyperlink{litobjrefs}{^{5}}$ & M6::$\gamma$$\hyperlink{litobjrefs}{^{6}}$ & $2523\pm40$\\
    6246004053326362368 & 16 17 43 & -18 58 18.3 & $16.7\pm0.5$ & 2MASS J16174255$-$1858179$\hyperlink{litobjrefs}{^{9}}$ & s/sdM7$\hyperlink{litobjrefs}{^{9}}$ & $2350\pm224$\\
    152917298349085824 & 4 25 16 & +28 29 27.1 & $7.2\pm0.1$ & 2MASS J04251550$+$2829275$\hyperlink{litobjrefs}{^{10}}$ & M7$\gamma$$\hyperlink{litobjrefs}{^{2}}$ & $2628\pm8$\\
    4364702279101281024 & 17 12 51 & -5 07 36.8 & $43.5\pm0.1$ & G  19$-$16B$\hyperlink{litobjrefs}{^{11}}$ & M7$\beta$$\hyperlink{litobjrefs}{^{12}}$ & $2410\pm55$\\
    6246979972975055360 & 15 57 52 & -19 56 39.5 & $19.9\pm0.4$ & UScoCTIO 135$\hyperlink{litobjrefs}{^{13}}$ & d/sdM7$\hyperlink{litobjrefs}{^{9}}$ & $2391\pm37$\\
    2497288672467622912 & 2 50 12 & -1 51 30.4 & $19.7\pm0.1$ & TVLM 831$-$154910$\hyperlink{litobjrefs}{^{14}}$ & M7.3$\gamma$$\hyperlink{litobjrefs}{^{6}}$ & $2664\pm20$\\
    638128236336998016 & 9 24 31 & +21 43 51.9 & $9.9\pm0.5$ & 2MASS J09243114$+$2143536$\hyperlink{litobjrefs}{^{15}}$ & M7$\beta$$\hyperlink{litobjrefs}{^{15}}$ & $2534\pm61$\\
    5682841554856156160 & 9 17 11 & -16 50 05.3 & $13.7\pm0.3$ & SIPS J0917$-$1649$\hyperlink{litobjrefs}{^{16}}$ & M7$\beta$$\hyperlink{litobjrefs}{^{15}}$ & $2532\pm57$\\
    1191334936190541184 & 15 56 19 & +13 00 53.4 & $10.9\pm0.7$ & 2MASS J15561873$+$1300527$\hyperlink{litobjrefs}{^{17}}$ & M8$\beta$$\hyperlink{litobjrefs}{^{17}}$ & $2387\pm153$\\
    1250625276082413568 & 13 54 43 & +21 50 29.4 & $11.1\pm0.3$ & 2MASS J13544271$+$2150309$\hyperlink{litobjrefs}{^{15}}$ & M8$\gamma$$\hyperlink{litobjrefs}{^{15}}$ & $2593\pm49$\\
    1597899151767870208 & 15 41 24 & +54 25 58.7 & $7.8\pm0.4$ & 2MASS J15412408$+$5425598$\hyperlink{litobjrefs}{^{17}}$ & sdM7.5$\hyperlink{litobjrefs}{^{18}}$ & $2480\pm140$\\
    1310888340170379136 & 16 39 08 & +28 39 00.6 & $9.3\pm0.5$ & 2MASS J16390818$+$2839015$\hyperlink{litobjrefs}{^{17}}$ & M8$\beta$$\hyperlink{litobjrefs}{^{15}}$ & $2516\pm54$\\
    4562040220870331520 & 17 03 36 & +21 19 03.1 & $12.8\pm0.5$ & 2MASS J17033593$+$2119071$\hyperlink{litobjrefs}{^{15}}$ & M8$\beta$$\hyperlink{litobjrefs}{^{15}}$ & $2416\pm135$\\
    6442586188225229312 & 20 11 57 & -62 01 18.9 & $12.8\pm0.4$ & 2MASS J20115649$-$6201127$\hyperlink{litobjrefs}{^{19}}$ & sdM8$\hyperlink{litobjrefs}{^{20}}$ & $2422\pm51$\\
    4588438567346043776 & 18 26 08 & +30 14 07.9 & $90.1\pm0.1$ & LSR J1826$+$3014$\hyperlink{litobjrefs}{^{21}}$ & sdM8.5$\hyperlink{litobjrefs}{^{18}}$ & $2360\pm14$\\
    147786354323787008 & 4 34 06 & +24 18 50.4 & $7.5\pm0.2$ & 2MASS J04340619$+$2418508$\hyperlink{litobjrefs}{^{22}}$ & M8$\gamma$$\hyperlink{litobjrefs}{^{2}}$ & $2440\pm67$\\
    1938820873903912448 & 23 36 38 & +45 23 30.4 & $8.0\pm0.7$ & 2MASS J23363834$+$4523306$\hyperlink{litobjrefs}{^{17}}$ & M8$\beta$$\hyperlink{litobjrefs}{^{17}}$ & $2531\pm83$\\
    4693823801926111360 & 2 21 29 & -68 31 40.1 & $14.4\pm0.2$ & 2MASS J02212859$-$6831400$\hyperlink{litobjrefs}{^{23}}$ & M8$\hyperlink{litobjrefs}{^{23}}$ & $2471\pm63$\\
    4708433867622492416 & 0 38 15 & -64 03 53.7 & $21.8\pm0.3$ & 2MASS J00381489$-$6403529$\hyperlink{litobjrefs}{^{5}}$ & M8.2$\beta$$\hyperlink{litobjrefs}{^{6}}$ & $2252\pm63$\\
    5734132118729087488 & 8 56 14 & -13 42 24.6 & $18.6\pm0.2$ & 2MASS J08561384$-$1342242$\hyperlink{litobjrefs}{^{6}}$ & M8.6$\beta$$\hyperlink{litobjrefs}{^{6}}$ & $2380\pm32$\\
    6258149537937551232 & 15 20 17 & -17 55 34.5 & $21.5\pm0.3$ & SIPS J1520$-$1755$\hyperlink{litobjrefs}{^{16}}$ & M8$\beta$$\hyperlink{litobjrefs}{^{15}}$ & $2353\pm63$\\
    4815936868977501568 & 4 36 28 & -41 14 46.3 & $25.3\pm0.1$ & 2MASS J04362788$-$4114465$\hyperlink{litobjrefs}{^{24}}$ & M8$\beta$$\gamma$$\hyperlink{litobjrefs}{^{25}}$ & $2429\pm15$\\
    373562923829421440 & 1 14 58 & +43 18 57.6 & $21.1\pm0.4$ & 2MASS J01145788$+$4318561$\hyperlink{litobjrefs}{^{26}}$ & M8$\beta$$\hyperlink{litobjrefs}{^{26}}$ & $2213\pm102$\\
    5203361404618057984 & 9 45 14 & -77 53 14.0 & $15.4\pm0.1$ & 2MASS J09451445$-$7753150$\hyperlink{litobjrefs}{^{6}}$ & M8.2$\beta$$\hyperlink{litobjrefs}{^{6}}$ & $2425\pm20$\\
    6407490636060550400 & 22 35 36 & -59 06 32.0 & $21.3\pm0.2$ & 2MASS J22353560$-$5906306$\hyperlink{litobjrefs}{^{5}}$ & M8.6$\beta$$\hyperlink{litobjrefs}{^{6}}$ & $2289\pm80$\\
    1349492949336359936 & 17 50 13 & +44 24 06.7 & $32.5\pm0.3$ & LSPM J1750$+$4424$\hyperlink{litobjrefs}{^{27}}$ & M8$\beta$$\hyperlink{litobjrefs}{^{28}}$ & $2525\pm26$\\
    6468916639853825664 & 20 28 22 & -56 37 03.5 & $15.2\pm0.2$ & 2MASS J20282203$-$5637024$\hyperlink{litobjrefs}{^{5}}$ & M8$\gamma$$\hyperlink{litobjrefs}{^{6}}$ & $2417\pm41$\\
    553593388644803968 & 5 38 17 & +79 31 05.4 & $43.1\pm0.0$ & LP   16$-$36$\hyperlink{litobjrefs}{^{29}}$ & sdM$\hyperlink{litobjrefs}{^{29}}$ & $2671\pm10$\\
    6568517687360642816 & 22 22 56 & -44 46 22.5 & $21.3\pm0.3$ & SIPS J2222$-$4446$\hyperlink{litobjrefs}{^{16}}$ & M8$\beta$$\hyperlink{litobjrefs}{^{15}}$ & $2383\pm59$\\
    6551233295852532096 & 23 36 07 & -35 41 50.5 & $21.7\pm0.5$ & SIPS J2336$-$3541$\hyperlink{litobjrefs}{^{16}}$ & M8.6$\gamma$$\hyperlink{litobjrefs}{^{6}}$ & $2268\pm66$\\
    5401822669314874240 & 11 02 10 & -34 30 35.8 & $16.9\pm0.1$ & TWA 28$\hyperlink{litobjrefs}{^{30}}$ & M8.5$\gamma$$\hyperlink{litobjrefs}{^{31}}$ & $2382\pm42$\\
    2861861847492765568 & 0 08 28 & +31 25 58.0 & $11.4\pm0.6$ & 2MASS J00082822$+$3125581$\hyperlink{litobjrefs}{^{26}}$ & M8$\gamma$$\hyperlink{litobjrefs}{^{26}}$ & $2292\pm203$\\
    5657734928392398976 & 9 38 40 & -27 48 21.2 & $35.3\pm0.1$ & SIPS J0938$-$2748$\hyperlink{litobjrefs}{^{16}}$ & M8$\beta$$\hyperlink{litobjrefs}{^{15}}$ & $2476\pm11$\\
    656167618671591424 & 8 19 46 & +16 58 53.3 & $33.0\pm0.3$ & 2MASS J08194602$+$1658539$\hyperlink{litobjrefs}{^{32}}$ & M8$\beta$$\hyperlink{litobjrefs}{^{18}}$ & $2350\pm43$\\
    5432903251692290944 & 9 39 59 & -38 17 18.1 & $16.4\pm0.3$ & 2MASS J09395909$-$3817217$\hyperlink{litobjrefs}{^{15}}$ & M8$\gamma$$\hyperlink{litobjrefs}{^{15}}$ & $2406\pm34$\\
    147614422487144960 & 4 36 33 & +24 21 39.4 & $6.3\pm0.1$ & 2MASS J04363248$+$2421395$\hyperlink{litobjrefs}{^{1}}$ & M8$\gamma$$\hyperlink{litobjrefs}{^{2}}$ & $2457\pm11$\\
    3313381382679891456 & 4 32 51 & +17 30 08.9 & $6.9\pm0.4$ & 2MASS J04325119$+$1730092$\hyperlink{litobjrefs}{^{33}}$ & M8$\gamma$$\hyperlink{litobjrefs}{^{34}}$ & $2373\pm67$\\
    1952664279346269056 & 21 40 39 & +36 55 55.3 & $9.9\pm0.4$ & 2MASS J21403907$+$3655563$\hyperlink{litobjrefs}{^{15}}$ & M8$\beta$$\hyperlink{litobjrefs}{^{15}}$ & $2517\pm42$\\
    3459372646830687104 & 12 07 33 & -39 32 54.4 & $15.5\pm0.1$ & TWA 27$\hyperlink{litobjrefs}{^{35}}$ & M8$\beta$$\hyperlink{litobjrefs}{^{36}}$ & $2430\pm13$\\
    3459725624422311424 & 12 03 59 & -38 21 40.6 & $12.2\pm0.2$ & TWA 38$\hyperlink{litobjrefs}{^{5}}$ & M8$\gamma$$\hyperlink{litobjrefs}{^{31}}$ & $2455\pm22$\\
    6281432246412503424 & 14 44 17 & -20 19 56.9 & $58.1\pm0.1$ & SSSPM J1444$-$2019$\hyperlink{litobjrefs}{^{37}}$ & sdM9$\hyperlink{litobjrefs}{^{38}}$ & $2352\pm10$\\
    5399990638128330752 & 11 06 45 & -37 15 11.7 & $9.8\pm0.3$ & 2MASS J11064461$-$3715115$\hyperlink{litobjrefs}{^{5}}$ & M9.4$\gamma$$\hyperlink{litobjrefs}{^{6}}$ & $2396\pm65$\\
    2898019875782441856 & 6 08 53 & -27 53 58.2 & $22.6\pm0.2$ & DENIS J060852.8$-$275358$\hyperlink{litobjrefs}{^{32}}$ & M9$\beta$$\hyperlink{litobjrefs}{^{25}}$ & $2359\pm102$\\
    216704503361774080 & 3 45 21 & +32 18 17.6 & $3.1\pm0.1$ & 2MASS J03452106$+$3218178$\hyperlink{litobjrefs}{^{39}}$ & M9$\gamma$$\hyperlink{litobjrefs}{^{40}}$ & $2588\pm12$\\
    6152893526035165312 & 12 47 44 & -38 16 46.8 & $11.9\pm0.3$ & 2MASS J12474428$-$3816464$\hyperlink{litobjrefs}{^{41}}$ & M9$\hyperlink{litobjrefs}{^{6}}$ & $2380\pm98$\\
    6236753694496012544 & 15 47 47 & -24 23 51.7 & $29.3\pm0.3$ & DENIS J154747.2$-$242349$\hyperlink{litobjrefs}{^{23}}$ & L0$\beta$$\hyperlink{litobjrefs}{^{36}}$ & $2273\pm74$\\
    6358389917097619968 & 21 54 49 & -74 59 14.9 & $21.3\pm0.2$ & 2MASS J21544859$-$7459134$\hyperlink{litobjrefs}{^{5}}$ & M9.8$\gamma$$\hyperlink{litobjrefs}{^{6}}$ & $2325\pm32$\\
    6366726276822544768 & 20 00 49 & -75 23 08.8 & $34.0\pm0.1$ & SIPS J2000$-$7523$\hyperlink{litobjrefs}{^{42}}$ & M9$\gamma$$\hyperlink{litobjrefs}{^{43}}$ & $2338\pm32$\\
    365582359196918656 & 0 41 22 & +35 47 12.5 & $9.3\pm1.1$ & 2MASS J00412179$+$3547133$\hyperlink{litobjrefs}{^{17}}$ & sdM9$\hyperlink{litobjrefs}{^{44}}$ & $2194\pm145$\\
    2969695320811729280 & 5 26 43 & -18 24 31.9 & $18.6\pm0.1$ & 2MASS J05264316$-$1824315$\hyperlink{litobjrefs}{^{5}}$ & M6.2$\gamma$$\hyperlink{litobjrefs}{^{6}}$ & $2663\pm12$\\
    6845967936118138752 & 20 13 52 & -28 06 03.3 & $21.0\pm0.3$ & 2MASS J20135152$-$2806020$\hyperlink{litobjrefs}{^{23}}$ & L0$\beta$$\hyperlink{litobjrefs}{^{36}}$ & $2277\pm68$\\
    3230008650057256960 & 4 43 38 & +0 02 03.4 & $47.6\pm0.1$ & 2MASSI J0443376$+$000205$\hyperlink{litobjrefs}{^{45}}$ & M9$\beta$$\hyperlink{litobjrefs}{^{46}}$ & $2290\pm35$\\
    6096164227899898880 & 14 11 42 & -45 24 20.1 & $19.1\pm0.2$ & 2MASS J14114474$-$4524153$\hyperlink{litobjrefs}{^{47}}$ & sdM9$\hyperlink{litobjrefs}{^{48}}$ & $2487\pm47$\\
    3478519134297202560 & 11 39 51 & -31 59 21.8 & $21.4\pm0.2$ & TWA 26$\hyperlink{litobjrefs}{^{35}}$ & M9$\gamma$$\hyperlink{litobjrefs}{^{35}}$ & $2390\pm17$\\
    1320853355787534848 & 15 52 59 & +29 48 47.5 & $48.9\pm0.2$ & 2MASS J15525906$+$2948485$\hyperlink{litobjrefs}{^{49}}$ & L0$\gamma$$\hyperlink{litobjrefs}{^{50}}$ & $2097\pm49$\\
    6132672029732817024 & 12 45 14 & -44 29 08.1 & $12.2\pm0.3$ & TWA 29$\hyperlink{litobjrefs}{^{51}}$ & L0$\gamma$$\hyperlink{litobjrefs}{^{36}}$ & $2317\pm41$\\
    1458522725665649536 & 13 47 50 & +33 36 01.5 & $13.0\pm0.7$ & 2MASS J13474972$+$3336019$\hyperlink{litobjrefs}{^{52}}$ & sdL0$\hyperlink{litobjrefs}{^{53}}$ & $2387\pm70$\\
    4568719543555702272 & 17 11 13 & +23 26 32.5 & $30.9\pm0.3$ & 2MASSI J1711135$+$232633$\hyperlink{litobjrefs}{^{46}}$ & L1$\gamma$$\hyperlink{litobjrefs}{^{36}}$ & $2065\pm90$\\
    2328674716056981888 & 23 22 47 & -31 33 32.1 & $50.2\pm0.2$ & 2MASS J23224684$-$3133231$\hyperlink{litobjrefs}{^{23}}$ & L0$\gamma$$\hyperlink{litobjrefs}{^{23}}$ & $2017\pm46$\\
    144711230753602048 & 4 35 36 & +21 15 03.6 & $16.7\pm0.6$ & 2MASS J04353511$+$2115201$\hyperlink{litobjrefs}{^{47}}$ & sdL0$\hyperlink{litobjrefs}{^{54}}$ & $2371\pm74$\\
    5183457632811832960 & 3 06 02 & -3 31 06.1 & $24.7\pm0.3$ & 2MASS J03060140$-$0330438$\hyperlink{litobjrefs}{^{47}}$ & sdL0$\hyperlink{litobjrefs}{^{54}}$ & $2348\pm55$\\
    4954323704550180352 & 1 41 58 & -46 33 58.1 & $27.3\pm0.4$ & 2MASS J01415823$-$4633574$\hyperlink{litobjrefs}{^{55}}$ & L0 $\gamma$$\hyperlink{litobjrefs}{^{25}}$ & $2146\pm153$\\
    4980384088633481216 & 0 32 56 & -44 05 07.3 & $29.0\pm0.4$ & EROS$-$MP J0032$-$4405$\hyperlink{litobjrefs}{^{56}}$ & L0 $\gamma$$\hyperlink{litobjrefs}{^{50}}$ & $2092\pm83$\\
    4841448081361281920 & 3 57 27 & -44 17 30.5 & $21.3\pm0.3$ & 2MASS J03572695$-$4417305$\hyperlink{litobjrefs}{^{57}}$ & L0 $\beta$$\hyperlink{litobjrefs}{^{25}}$ & $2213\pm115$\\
    2358397882610264960 & 1 16 39 & -16 54 20.1 & $16.1\pm0.5$ & 2MASS J01163865$-$1654210$\hyperlink{litobjrefs}{^{58}}$ & sdL0$\hyperlink{litobjrefs}{^{53}}$ & $2291\pm96$\\
    2802623115925093760 & 0 43 26 & +22 21 21.9 & $15.0\pm0.3$ & 2MASS J00432610$+$2221295$\hyperlink{litobjrefs}{^{47}}$ & sdL1$\hyperlink{litobjrefs}{^{54}}$ & $2410\pm36$\\
    4584405146372926720 & 17 56 10 & +28 15 16.8 & $28.9\pm0.3$ & 2MASS J17561080$+$2815238$\hyperlink{litobjrefs}{^{15}}$ & sdL1$\hyperlink{litobjrefs}{^{15}}$ & $2032\pm108$\\
    1047188004010109440 & 10 22 47 & +58 25 33.6 & $54.0\pm0.2$ & 2MASS J10224821$+$5825453$\hyperlink{litobjrefs}{^{59}}$ & L1$\gamma$$\hyperlink{litobjrefs}{^{25}}$ & $2028\pm68$\\
    1060313492785021312 & 11 08 30 & +68 30 13.5 & $61.8\pm0.1$ & LSPM J1108$+$6830$\hyperlink{litobjrefs}{^{27}}$ & L1{$\gamma$}$\hyperlink{litobjrefs}{^{6}}$ & $2019\pm55$\\
    2955015805492793088 & 5 18 46 & -27 56 45.8 & $18.3\pm0.6$ & 2MASSI J0518461$-$275645$\hyperlink{litobjrefs}{^{46}}$ & L1$\beta$$\hyperlink{litobjrefs}{^{46}}$ & $2183\pm164$\\
    2781513733917711616 & 0 45 22 & +16 34 44.0 & $65.4\pm0.2$ & 2MASS J00452143$+$1634446$\hyperlink{litobjrefs}{^{60}}$ & L2 $\beta$$\hyperlink{litobjrefs}{^{50}}$ & $2018\pm39$\\
    824017070904063488 & 10 04 20 & +50 22 56.1 & $46.2\pm0.5$ & G 196$-$3B$\hyperlink{litobjrefs}{^{61}}$ & L3$\gamma$$\hyperlink{litobjrefs}{^{25}}$ & $1899\pm100$\\
    3303349202364648320 & 3 55 24 & +11 33 33.7 & $109.1\pm0.5$ & 2MASS J03552337$+$1133437$\hyperlink{litobjrefs}{^{62}}$ & L5 $\gamma$$\hyperlink{litobjrefs}{^{50}}$ & $1839\pm140$\\
\end{longtable}

\bsp	
\label{lastpage}
\end{document}